\documentclass[acmtocl]{acmtrans2m}

\title{A Simplified Suspension Calculus and its \\
Relationship to Other Explicit Substitution Calculi}

\author{Andrew Gacek and Gopalan Nadathur \\
Digital Technology Center and Department of Computer Science and
Engineering\\
University of Minnesota}

\def\firstfoot{\def\@firstfoot{}}
\def\runningfoot{\def\@runningfoot{}}

\def\permission{}

\begin{abstract}
This paper concerns the explicit treatment of substitutions in the
lambda calculus. One of its contributions is the simplification and
rationalization of the suspension calculus that embodies such a
treatment. The earlier version of this calculus provides a cumbersome
encoding of substitution composition, an operation that is important
to the efficient realization of reduction. This encoding is simplified
here, resulting in a treatment that is easy to use directly in
applications. The rationalization consists of the elimination of a
practically inconsequential flexibility in the unravelling of
substitutions that has the inadvertent side effect of losing
contextual information in terms; the modified calculus now has a
structure that naturally supports logical analyses, such as ones
related to the assignment of types, over lambda terms. The overall
calculus is shown to have pleasing theoretical properties such as a
strongly terminating sub-calculus for substitution and confluence even
in the presence of term meta variables that are accorded a grafting
interpretation. Another contribution of the paper is the
identification of a broad set of properties that are desirable for
explicit substitution calculi to support and a classification of a
variety of proposed systems based on these. The suspension calculus is
used as a tool in this study. In particular, mappings are described
between it and the other calculi towards understanding the
characteristics of the latter.
\end{abstract}


\usepackage{amsmath}
\usepackage{amssymb}
\usepackage{proof}
\usepackage{url}

\long\def\ignore#1{}

\newcommand{\ra}{\rightarrow}

\newcommand{\app}{{\ }}

\newcommand{\lambdadb}{\lambda \,}

\newcommand{\dum}[1]{@ #1}

\newcommand{\lenv}{{\lbrack\!\lbrack}}
\newcommand{\renv}{{\rbrack\!\rbrack}}
\newcommand{\env}[1]{{\lenv #1 \renv}}

\newcommand{\lmenv}{{\lbrace\!\!\lbrace}}
\newcommand{\rmenv}{{\rbrace\!\!\rbrace}}
\newcommand{\menv}[1]{{\lmenv #1 \rmenv}}

\newcommand{\lmenvt}{{\langle\!\langle}}
\newcommand{\rmenvt}{{\rangle\!\rangle}}
\newcommand{\menvt}[1]{{\lmenvt #1 \rmenvt}}

\newcommand{\monussign}{{\stackrel{.}{\;\overline{\,\,\,}\;}}}
\newcommand{\monus}[2]{{#1 \monussign #2}}

\newcommand{\one}[1]{\rhd_{\!#1}}
\newcommand{\oneread}{{\rhd\!_{r}}}
\newcommand{\onemerge}{{\rhd\!_{m}}}
\newcommand{\onebetas}{{\rhd\!_{\beta_s}}}
\newcommand{\onebetaspar}{{\rhd\!_{\beta_s\parallel}}}
\newcommand{\onereadmerge}{{\rhd\!_{rm}}}
\newcommand{\onereadbetas}{{\rhd\!_{r\beta_s}}}
\newcommand{\oneall}{\rhd\!_{rm \beta_s}}

\newcommand{\readmerge}{{\rhd\!_{rm}^*}}
\newcommand{\readbetas}{{\rhd\!_{r\beta_s}^* }}
\newcommand{\betas}{{\rhd\!_{\beta_s}^*}}
\newcommand{\readmergebetas}{\rhd\!_{rm \beta_s}^*}

\newcommand{\many}[1]{\rhd_{\!#1}^*}

\newcommand{\onebeta}{{\rhd\!_{\beta}}}
\newcommand{\betared}{{\rhd\!_{\beta}^* }}

\newcommand{\rnnorm}[1]{{\vert #1 \vert}}

\newcommand{\ess}{{\cal E}}

\newcommand{\ie}{i.e.}
\newcommand{\eg}{e.g.}

\newcounter{itemno}

\newtheorem{theorem}{Theorem}[section]
\newtheorem{lemma}[theorem]{Lemma}

\newdef{defn}[theorem]{Definition}

\newcommand{\lift}{\mathop{\Uparrow}}

\newcommand{\sig}[1]{\mathop{\sigma^{#1}}}
\newcommand{\ph}[2]{\mathop{\varphi_{#1}^{#2}}}

\begin{document}
\begin{bottomstuff}
Authors' addresses: \\
A. Gacek, University of Minnesota, 4-192 EE/CS Building, 200 Union
Street SE, Minneapolis, MN 55455, USA, Email:
\verb+andrew.gacek@gmail.com+\\
G. Nadathur, University of Minnesota, 4-192 EE/CS
Building, 200 Union Street SE, Minneapolis, MN 55455, USA, Email:
\verb+gopalan@cs.umn.edu+
\end{bottomstuff}

\maketitle

\section{Introduction}

This paper concerns the explicit treatment of substitution in the
lambda calculus. It has a twofold purpose within this context. First,
it simplifies and rationalizes a particular calculus known as the {\it
  suspension calculus} that provides such a treatment
\cite{NW98tcs}. Second, using the resulting system as a basis, it 
attempts to explicate the nuances of and differences between an array
of explicit substitution calculi that have been proposed in recent 
years. 

The desire to treat substitution directly in the syntax and rewrite
rules of the lambda calculus has had a variety of motivations. The
suspension calculus was developed originally with the intention of
supporting a higher-order view of syntax, now commonly referred to as
{\it higher-order abstract syntax} \cite{PE88pldi} or {\it lambda tree
syntax} \cite{Mil00cl}. Success has been encountered in this
endeavour: amongst other applications, the notation has been employed
in the reasoning system called {\it Bedwyr} \cite{BGMNT07}, in the
abstract machine for $\lambda$Prolog \cite{NM99cade} and in the 
implementation of the {\it FLINT} typed intermediate
language \cite{shao98:imp}. Despite its use in practical systems, the
original suspension calculus manifests some deficiencies. One problem
is the building in of excessive flexibility in the unravelling of
substitutions that leads inadvertently to the loss of certain kinds of
context information. This added flexibility does not really enhance
the efficiency of reduction and has unpleasant side effects such as
the loss of the ability to associate a typing calculus with lambda
terms. Another problem relates to the encoding of the composition of
substitutions. Although the notation includes such a capability, its
treatment is complicated and has led to the description of a derived
calculus \cite{Nad99jflp} that is the one usually employed  
in applications. A drawback with this derived calculus is that it
does not possess the property of confluence when meta variables are
added to the syntax under the so-called {\it grafting}  
interpretation\footnote{Although this has not been
made explicit previously, the original suspension calculus is
confluent even in the presence of graftable meta variables.}. At a
practical level, this has the impact that new approaches to
higher-order unification based on using graftable meta variables
\cite{DHK00} cannot be exploited relative to it.

One contribution of this paper is the redressing of this situation. In 
particular, it describes a modified treatment of substitution
composition that is simultaneously natural, easy to use directly in
implementations and consistent with contextual properties. 

The last fifteen years has seen the description of a large number of
explicit substitution calculi, often without a clear enunciation of
the goals underlying their design. A consequence of this phenomenon is
that it has been difficult to evaluate the different calculi or even
to understand the distinctive characteristics of each. This paper
contributes in a second way by bringing greater clarity to these
matters.  Specifically, it identifies three properties that appear 
important for explicit treatments of substitution to support. It then
surveys some of the prominent calculi in this realm through this 
prism. The suspension calculus that is developed in the earlier
sections serves as a tool in understanding the various other
systems. Through this process, a better grasp is also obtained of the
capabilities of this specific notation.

The rest of the paper is structured as follows. In the next section we
describe the new version of the suspension
calculus. Section~\ref{sec:properties} then elucidates its properties:
we show here the strong normalizability and confluence of the
sub-calculus for treating substitutions and the confluence of the
overall calculus even in the presence of graftable meta
variables. Section~\ref{sec:compare} discusses other treatments of
explicit substitutions and contrasts these with the one developed
here. Section~\ref{sec:conc} concludes the paper.

\section{The Suspension Calculus}\label{sec:notation}

The modified version of the suspension calculus of Nadathur and Wilson
\citeyear{NW98tcs} that we present in this section does not sacrifice
any of the computational properties of the original calculus that are
essential to its use in implementations. Rather, it embodies a view of
it that is easier to reason about and to relate to other approaches to
explicit substitutions.  In the first two subsections below, we
outline the intuitions underlying the suspension calculus and then
substantiate this discussion through a precise description of its
syntax and reduction rules. We then discuss the relationship of the
version of the calculus we present here with the original version and
also describe variants of it arising from the introduction of meta
variables under two different interpretations.

\subsection{Motivating the Encoding of Substitutions}\label{ssec:motivation}

We are interested in enhancing the syntax of the lambda calculus with
a new category of expressions that is capable of encoding terms
together with substitutions that have yet to be carried out on
them. The kinds of substitutions that we wish to treat are those that
arise from beta contraction steps being applied to lambda terms. 
Towards understanding what needs to be encoded in this context, we may
consider a term with the following structure:
\begin{tabbing}
\qquad\=\kill
\>$(\ldots ((\lambdadb \ldots(\lambdadb \ldots ((\lambdadb \ldots t
\ldots)\app s_1)\ldots )\ldots)\app s_2) \ldots)$
\end{tabbing}
We assume here a de Bruijn representation for lambda terms, \ie, names
are not used with abstractions and bound variable occurrences are
replaced by indices that count abstractions back up to the one
binding them \cite{debruijn72}.
We have elided much of the detail in the term shown and have, in fact,
focussed only on the following aspects: there is a beta redex in it (whose
``argument'' part is $s_2$) that is embedded possibly under
abstractions and that itself contains at least another embedded beta
redex. Contracting the two beta redexes shown should produce a term of
the form 
\begin{tabbing}
\qquad\=\kill
\>$(\ldots (\ldots(\lambdadb \ldots ( \ldots t'
\ldots)\ldots )\ldots) \ldots)$
\end{tabbing}
where $t'$ is obtained from $t$ by substituting $s_2$ and (a modified
form of) $s_1$ for appropriate variables and adjusting the indices for
other bound variables to account for the disappearance of two
enclosing abstractions. Our goal is to represent $t'$ as $t$ coupled
with the substitutions that are to be performed on it. 

Towards developing a suitable encoding, it is useful to
factor the variable references within $t$ into two groups: those that
are bound by abstractions inside the first beta redex that is
contracted and those that are bound by abstractions enclosing this
redex. Let us refer to the number of abstractions enclosing a term in
a particular context as its embedding level relative to that
context. For example, if we assume that every abstraction within the
outer beta redex in the term considered above has been explicitly
shown, then the embedding level of $t$ in this 
context is $3$. Rewriting a beta redex eliminates abstractions and
therefore changes embedding levels. Thus, if the two beta redexes of
interest are both contracted, the embedding level of $t$ becomes
$1$. We shall call the embedding levels at a term before and after
beta contractions the {\it old} and {\it new} embedding levels
respectively. Simply recording these with a term is enough for
encoding the change that needs to be made to the indices for variables
bound by the ``outer'' group of abstractions; in particular, these
indices must be decreased by the difference between the old and the
new embedding levels.

Substitutions for the other group of variable references, \ie, those
bound by abstractions within the first beta redex contracted, can be
recorded explicitly in an environment. To suggest a concrete syntax,
the term $t'$ in the example considered may be represented by the
expression $\env{t,ol,nl,e}$ where $ol$ and $nl$ are the old and new
embedding levels, respectively, and $e$ is the environment. Note that
the number of entries in the environment must coincide with the old
embedding level. It is convenient also to maintain the environment as
a list or sequence of elements whose order is reverse that of the
embedding level of the abstraction they correspond to; amongst other
things, this allowed for an easy augmentation of the environment in a
top-down traversal of the term. Now, one component of the entry for an
abstraction that is contracted should obviously be the argument part
of the relevant beta redex. For an abstraction not eliminated by a
contraction, there is no new term to be substituted, but we can still
correctly record the index corresponding to the first free variable as
a pseudo substitution for it. In both these cases, we have also to pay
attention to the following fact: the term in the environment may be
substituted into a new context that has a larger number of enclosing
abstractions and hence de Bruijn indices for free variables within it
may have to be modified. To encode this renumbering, it suffices to
record the (new) embedding level at the relevant abstraction with the
environment entry. The difference between this and the (new) embedding
level at the point of substitution determines the amount by which the
free variable indices inside the term being substituted have to be
changed. Thus, each environment entry has the form $(t,l)$ where $t$
is a term and $l$ is a positive number. We refer to the second
component of each such entry as its index and we observe that the
indices for successive environment entries must form a non-increasing
sequence at least for the simple form of environments we are presently
considering.

Once we have permitted terms encoding substitutions into our syntax,
it is possible for such terms to appear one inside another. A
particular instance of this phenomenon is when they appear in
juxtaposition as in the term
\begin{tabbing}
\qquad\=\kill
\>$\env{\env{t,ol_1,nl_1,e_1},ol_2,nl_2,e_2}$.
\end{tabbing}
This term
corresponds to separately performing two sets of substitutions into
$t$. It is useful to have a means for combining these into one set of
substitutions, \ie, for rewriting the indicated
term into one of the form $\env{t,ol',nl',e'}$. 
In determining the shape of the new term, it is useful to note that 
$e_1$ and $e_2$ represent substitutions for overlapping
sequences of abstractions within which $t$ is embedded. 
The generation of the original term can, in fact, be visualized as
follows: 
First, a walk is made over $ol_1$ abstractions immediately
enclosing $t$, possibly eliminating some of them via beta
contractions, recording substitutions for all of them in $e_1$ and
eventually leaving behind $nl_1$ enclosing abstractions. 
Then a similar walk is made over $ol_2$ abstractions immediately
enclosing the term $\env{t_1,ol_1,nl_1,e_1}$, recording 
substitutions for each of them in $e_2$ and leaving behind $nl_2$
abstractions. 
Notice that the $ol_2$ 
abstractions scanned in the second walk are coextensive with some
final segment of the $nl_1$ abstractions left behind after the first
walk and includes additional abstractions if $ol_2 > nl_1$. 

Based on the image just evoked, it is not difficult to see
what $ol'$ in the term representing the combined form for the
substitutions should be: this form represents a walk over $ol_1$
enclosing abstractions in the case that $ol_2 
\leq nl_1$ and $ol_1 + (ol_2 - nl_1)$ abstractions otherwise and
$ol'$ should be the appropriate one of these values. 
Similarly, the number of abstractions eventually left behind is $nl_2$
or $nl_2 + (nl_1 - ol_2)$ depending on whether or not $nl_1 \leq
ol_2$, and this determines the value of $nl'$. With regard to the
environment $e'$, this should be composed of the elements of $e_1$
modified by the substitutions encoded in $e_2$ followed by a final
segment of $e_2$ in the case that $ol_2 > nl_1$. The modification to
be effected on the elements of $e_1$ may be understood as
follows. Suppose $e_1$ has as an element the pair $(s,l)$. Then $s$ is
affected by only that part of $e_2$ that comes after the first $nl_1 -
l$ entries in it. Further, the index of the corresponding entry in the
composite environment would have to be increased from $l$ by an amount
equal to $ol_2 - nl_1$ in the case that $ol_2 > nl_1$. From these
observations, it is clear that the merged environment can be generated
completely from the components $e_1$, $nl_1$, $ol_2$ and $e_2$. We
correspondingly choose to encode this environment by the expression
$\menv{e_1,nl_1,ol_2,e_2}$.  

Our focus here has been on motivating the new syntactic forms in the
suspension calculus. However, implicit in this discussion has been a
``meaning'' for these new expressions in the sense of a translation
into an underlying de Bruijn term. This informal semantics will be
made precise in the next section through a collection of rewrite rules
that can be used to incrementally ``calculate'' the intended encodings.

\subsection{The Syntax of Terms and the Rewriting System}\label{ssec:syntax}

We now describe precisely the collections of expressions that
constitute terms and environments in the suspension calculus. We
assume that the lambda terms to be treated contain constant symbols
drawn from a predetermined set. Letting $c$ represent such constants,
the $t$ and $e$ expressions given by the following rules
define a ``pre-syntax'' for our terms and environments: 
\begin{tabbing}
\qquad\=$t$\=\quad::=\quad\=\kill
\>$t$\>\quad::=\quad\>$c \ \vert \ \#i\ \vert\ (t\app t)\ \vert\
(\lambdadb t)\ \vert\ \env{t, n, n, e}$\\
\>$e$\> \quad::=\quad \> $nil\ \vert\ ((t,n) :: e)\ \vert\ \menv{e, n,
n, e}$
\end{tabbing}
In these rules, $n$ corresponds to the category of natural numbers and
$i$ represents positive integers. Terms of the form $(t_1\app
t_2)$ and $(\lambdadb t)$ are, as usual, referred to as applications
and abstractions. A term of the form $\#i$, known as a de Bruijn index,
represents a variable bound by the $i$th abstraction looking outward
from the point of its occurrence. Expressions of the form
$\env{t,ol,nl,e}$ are called {\it suspensions}; these constitute a
genuine extension to the syntax of lambda terms. The operator $::$ 
provides the means for forming lists in environments. We use the 
conventions that application is left associative, that $::$ is right
associative and that application binds more tightly than abstraction
to often omit parentheses in the expressions we write. We shall
sometimes need to suppress the distinction between terms and
environments and at these times we shall refer to them collectively as  
suspension expressions or, more simply, as expressions. 

The reason we think of the rules above as defining only the
pre-syntax is that we expect suspension expressions to also 
satisfy certain well-formedness constraints. In order to enunciate 
these constraints precisely, we need to associate the notions of {\it
  length} and {\it level} with environments. We do this through the
following definitions. The symbol $\monussign$ used in these 
definitions denotes the subtraction operation on natural numbers. 

\begin{defn}
The length of an environment $e$ is denoted by $len(e)$ and is defined
by recursion on its structure as follows: 
\begin{enumerate}
\item $len(nil) = 0$
\item $len((t,l)::e) = 1 + len(e)$
\item $len(\menv{e_1, nl_1, ol_2, e_2}) = len(e_1) +
  (\monus{len(e_2)}{nl_1})$ 
\end{enumerate}
\end{defn}

\begin{defn}
The level of an environment $e$, denoted by $lev(e)$, is also given by
recursion as follows:
\begin{enumerate}
\item $lev(nil) = 0$
\item $lev((t,l)::e) = l$
\item $lev(\menv{e_1, nl_1, ol_2, e_2}) = lev(e_2) +
(\monus{nl_1}{ol_2})$
\end{enumerate}
\end{defn}

The legitimacy requirements that complement the syntax rules is now
explicated as follows:

\begin{defn}\label{def:wellformed}
A suspension expression is considered well-formed just in case the
following conditions hold of all its subexpressions:
\begin{enumerate}
\item If it is of the form $\env{t, ol, nl, e}$ then $len(e) = ol$ and
  $lev(e) \leq nl$.
\item If it is of the form $(t,l)::e$ then $l \geq lev(e)$.
\item If it is of the form $\menv{e_1, nl_1, ol_2, e_2}$ then $lev(e_1) \leq
nl_1$ and $len(e_2) = ol_2$.
\end{enumerate}
\end{defn}
We henceforth consider only well-formed suspension expressions. We
shall also sometimes restrict our attention to environments which have
the structure of a list of bindings. We identify this class of
environments below.

\begin{defn}
A simple environment is one of the form 
\begin{tabbing}
\qquad\=\kill
\>$(t_0,l_0) :: (t_1,l_1) :: \ldots :: (t_{n-1}, l_{n-1}) :: nil$
\end{tabbing}
where by an abuse of notation, we allow $n$ to be $0$, in which case
the environment in question is $nil$. For $0 \leq i < n$, we write
$e[i]$ to denote the environment element $(t_i,l_i)$ and $e\{i\}$ to
denote $(t_i,l_i) :: \ldots :: (t_{n-1},l_{n-1}) :: nil$, \ie, the
environment obtained from $e$ by removing its first $i$ elements.  We
extend the last notation by letting $e\{i\}$ denote $nil$ in the case
that $i \geq {\it len}(e)$ for any simple environment $e$.
\end{defn}

\begin{figure}[t]
\begin{tabbing}
\quad\=(r11)\quad\=\kill
\> ($\beta_s$)\>$((\lambdadb t_1)\app t_2) \ra \env{t_1, 1, 0, (t_2,0)
  :: nil}$.\\[7pt] 
\> (r1)\> $\env{c,ol,nl,e} \ra c$, provided $c$ is a constant.\\[5pt]
\> (r2)\>$\env{\#i,0,nl,nil} \ra \#j$, where $j = i + nl$.\\[5pt] 
\> (r3)\> $\env{\#1,ol,nl,(t,l) :: e} \ra \env{t,0,nl',nil}$, where
$nl' = nl - l$.\\[5pt] 
\> (r4)\> $\env{\#i, ol, nl, (t,l) :: e} \ra \env{\# i',
ol',nl,e},$\\
\>\> where $i' = i - 1$ and $ol' = ol -1$, provided $i > 1$.\\[5pt] 
\> (r5)\> $\env{(t_1\app t_2),ol,nl,e} \ra (\env{t_1,ol,nl,e}\app
\env{t_2,ol,nl,e})$.\\[5pt] 
\> (r6)\>$\env{(\lambdadb t), ol, nl, e} \ra (\lambdadb \env{t,
ol', nl', (\#1, nl') :: e})$,\\
\>\> where $ol' = ol + 1$ and $nl' = nl + 1$.\\[7pt]
\> (m1)\>$\env{\env{t,ol_1,nl_1,e_1},ol_2,nl_2,e_2} \ra
\env{t,ol',nl', \menv{e_1,nl_1,ol_2,e_2}}$,\\
\>\>where $ol' = ol_1 + (\monus{ol_2}{nl_1})$ and $nl' = nl_2 +
(\monus{nl_1}{ol_2})$.\\[5pt]
\> (m2)\>$\menv{e_1, nl_1, 0, nil} \ra e_1$.\\[5pt] 
\> (m3)\>$\menv{nil, 0, ol_2, e_2} \ra e_2$.\\[5pt] 
\> (m4)\>$ \menv{nil, nl_1, ol_2, (t,l)::e_2} \ra \menv{nil, nl_1',
  ol_2', e_2}$,\\
\>\> where $nl_1' = nl_1 - 1$ and $ol_2' = ol_2 - 1$, provided $nl_1
\geq 1$.\\[5pt]
\> (m5)\> $ \menv{(t,n) :: e_1, nl_1, ol_2, (s,l) ::
e_2} \ra \menv{(t,n) :: e_1, nl_1', ol_2', e_2}$,\\
\>\>where $nl_1' = nl_1 - 1$ and $ol_2' = ol_2 - 1$, provided $nl_1 >
n$.\\[5pt] 
\> (m6)\> $\menv{(t,n) :: e_1, n, ol_2, (s,l) :: e_2} \ra
(\env{t, ol_2, l, (s,l) :: e_2}, m) :: \menv{e_1, n, ol_2, (s,l) :: e_2}$,\\
\>\>where $m = l + (\monus{n}{ol_2})$.
\end{tabbing}
\caption{Rewrite Rules for the Suspension Calculus}\label{fig:susprules}
\end{figure}

The rewrite system associated with suspension expressions comprises
three kinds of rules: the beta contraction rule that generates
substitutions, the {\it reading rules} that distribute them
over term structure and the {\it merging rules} that
allow for the combination of substitutions generated by different beta
contractions into a composite one. These three categories correspond
to the rules in Figure~\ref{fig:susprules} labelled $(\beta_s)$,
(r1)-(r6) and (m1)-(m6), respectively. The application of several of
these rules depends on arithmetic calculations on embedding levels and
indices. We have been careful in the formal presentation to identify
such calculations through side conditions on the rules. However, in
the sequel, we will often assimilate such arithmetic operations into
the rewrite rule itself with the understanding that they are to be
``interpreted.'' Using this approach, rule (r6) may have been written
instead as
\begin{tabbing}
\qquad\=\kill
\>$\env{(\lambdadb t), ol, nl, e} \ra (\lambdadb
\env{t, ol + 1, nl + 1, (\#1, nl + 1) :: e})$.
\end{tabbing}

\begin{defn}\label{def:rewriterelns}
We say that a suspension expression $r$ is related to $s$ by a
$\beta_s$-contraction step, a reading step or a merging step if it is
the result of applying the $(\beta_s$) rule, one of the rules
(r1)-(r6) or one of the rules (m1)-(m6), respectively, at any
relevant subexpression of $s$. We denote these relations by writing $s
\onebetas r$, $s \oneread r$ and $s \onemerge r$, respectively. The
union of the relations $\oneread$ and $\onemerge$ will be denoted by
$\onereadmerge$, that of $\oneread$ and $\onebetas$ by $\onereadbetas$
and, finally, that of all three relations by $\oneall$. If $R$
corresponds to any of these relations, we shall 
write $R^*$ to denote its reflexive and transitive closure. 
\end{defn}

The following theorem shows that these various relations are well-defined.

\begin{theorem}\label{th:wellformedness}
The relations $\onebetas$, $\oneread$ and $\onemerge$, and, hence, any
combination of them, preserve well-formedness of suspension
expressions. 
\end{theorem}

\begin{proof}
A somewhat stronger property can be proved for the rewriting relations of
interest: (i)~they leave the length of an environment unchanged,
(ii)~they never increase the level of an environment, and  (iii)~they
preserve well-formedness. These facts are established simultaneously by
induction on the structure of suspension expressions. The base case is
verified by considering in turn each rewrite rule in
Figure~\ref{fig:susprules}. The argument is then completed by
considering each possibility for the structure of an expression and
using the induction hypothesis. The details are entirely
straightforward and hence omitted.
\end{proof}

We illustrate the rewrite rules by considering their use on 
the term 
\begin{tabbing}
\qquad\=\kill
\>$((\lambdadb (\lambdadb \lambdadb \#1\app \#2\app \#3)\app t_2)\app
t_3),$ 
\end{tabbing}
where $t_2$ and $t_3$ are arbitrary terms. 
We trace a $\oneall$-rewrite sequence for this term below:
\begin{tabbing}
\qquad\=\kill
\>$((\lambdadb (\lambdadb \lambdadb \#1\app \#2\app \#3)\app t_2)\app
t_3)$\\ 
\>$\quad \betas \env{\env{\lambdadb \#1\app \#2\app \#3,1,0,(t_2,0)::nil},
1,0,(t_3,0)::nil}$ \\
\>$\quad \onemerge \env{\lambdadb \#1\app \#2\app \#3,2,0,\menv{
(t_2,0)::nil,0,1,(t_3,0)::nil}}$\\
\>$\quad \onemerge \env{\lambdadb \#1\app \#2\app \#3,2,0,(\env{t_2,1,0,(t_3,0)::nil},0)
:: \menv{ nil, 0,1, (t_3,0)::nil }}$ \\
\>$\quad \onemerge \env{\lambdadb \#1\app \#2\app
\#3,2,0,(\env{t_2,1,0,(t_3,0)::nil},0) :: (t_3,0) :: nil}.$ 
\end{tabbing}
\noindent The last expression in this sequence is a term that
represents, roughly, the ``suspended'' simultaneous substitution of $t_2$,
modified by the substitution of $t_3$ for its first free variable, and
of $t_3$ for the first two free variables in $(\lambdadb \#1\app \#2
\app \# 3)$. This suspension has been produced by contracting the two
beta redexes in the original term and then using the merging rules to
combine the two separate substitutions that are so generated. The
combined environment can now be moved inside the abstraction, distributed
over the applications and partially ``evaluated'' using the reading rules
to yield 
\begin{tabbing}
\qquad\=\kill
\>$(\lambdadb\#1\app \env{\env{t_2,1,0,(t_3,0)::nil},0,1,nil})\app
\env{t_3,0,1,nil}))$.
\end{tabbing}
This term manifests a structure that may be thought of as a
generalization of head-normal forms to suspension
terms. By applying reading and merging rules in accordance with the
structure of $t_2$ and $t_3$, we may further transform it into a
head-normal form in the conventional sense.

The terms in the de Bruijn style presentation of the lambda calculus
are a subset of the terms in the suspension calculus. In particular,
they are exactly the terms in the present notation that do not contain
any suspensions. Given a rewrite relation $R$, we shall say, as usual,
that an expression is in $R$-normal form if it cannot be further
transformed by the rules defining $R$. It is easily seen then that a
suspension term is in de Bruijn form just in case it is in
$\onereadmerge$-normal form. 
We would, of course, be interested in knowing if any given
suspension expression can be transformed into a normal form of this 
kind. We answer this question in the affirmative in the next section
and subsequently relate the rewrite relations defined here with the usual
notion of beta reduction over de Bruijn terms.

\subsection{Relationship to the Original Suspension
  Calculus}\label{ssec:simplification}

The suspension calculus as we have described it here deviates from the
original presentation in \cite{NW98tcs} in a few different ways. One
distinction arises from the use in the earlier version of the calculus
of a special form for the environment item that results from
percolating a substitution under an abstraction. These items are
written as $\dum{n}$ where $n$ is a natural number. The rule (r6)
correspondingly has the form 
\begin{tabbing}
\qquad\=\kill
\>$\env{(\lambdadb t), ol, nl, e} \ra (\lambdadb \env{t,
ol+1, nl+1, \dum{nl} :: e})$
\end{tabbing}
in that setting. This form was introduced into the syntax and treated
in special ways by the rewrite rules in anticipation of an
implementation optimization. It is, however, inessential at a
theoretical level. In particular, the behaviour of a dummy environment
element of the form $\dum{n}$ can be completely circumscribed by
replacing it with $(\#1,n+1)$\footnote{It should be noted, though,
  that the parsimony of the latter form is complemented by the
  introduction of more (perhaps unnecessary) possibilities for
  rewriting that considerably complicate the proof of termination for
  the reading and merging rules.}. We assume the impact of this
observation below.

Suspension expressions in the present setting constitute a
subset of the expressions in the original calculus at a pre-syntax
level. However, the well-formedness condition when restricted to these
expressions is different in the two contexts. The earlier condition
has a form that is identical to the one in
Definition~\ref{def:wellformed} except that the requirement on the 
levels of environments is replaced by one on their {\it indices}, a
notion that is defined below. 
\begin{defn}
Given a natural number $i$, the $i$-th index of an environment $e$ is
denoted by $ind_{i}(e)$ and is defined as follows:
\begin{enumerate}
\item If $e$ is $nil$ then $ind_{i}(e) = 0$.
\item If $e$ is $(t,k) :: e'$
then $ind_{i}(e)$ is $k$ if $i = 0$ and $ind_{i-1}(e')$ otherwise.
\item If $e$ is $\menv {e_1,nl,ol,e_2 }$, let
 $m = (\monus{nl}{ind_{i}(e_1)})$\footnote{The $\monussign$ here can be
 replaced by $-$ for well-formed expressions.} and $l = len(e_1)$. Then
\begin{tabbing}
\qquad\=\kill
\>$ind_{i}(e) = \left\{ \begin{array}{ll}
      ind_{m}(e_2) + (\monus{nl}{ol}) &
                \mbox{ if $i < l$ and  $len(e_2) > m$}\\ 
      ind_{i}(e_1)   & \mbox{ if $i < l$ and  $len(e_2) \leq  m$}\\
      ind_{(i - l+nl)}(e_2) & \mbox{ if $i \geq l$.}
                        \end{array} 
               \right.$
\end{tabbing}
\end{enumerate}
The index of an environment, denoted by $ind(e)$, is
$ind_{0}(e)$. 
\end{defn}
Any given environment expression $e$ is expected to be reducible to a
simple one of the form $(t_0,l_0) :: \ldots :: (t_{n-1},l_{n-1}) ::
nil$.  The $i$-th index of $e$ is then precisely $l_i$ if $i < n$
and $0$ otherwise. The level of $e$, in contrast, only {\it estimates}
the $0$-th index when $e$ is reduced to this simple form while
retaining information that is needed for interpreting intermediate
expressions in the rewriting process.  Nevertheless, we can observe
the following:

\begin{lemma}
The well-formed expressions of the suspension calculus as described in 
this paper are a subset of the well-formed ones of the original
presentation. 
\end{lemma}

\begin{proof} We prove the following by induction on
  the structure of a suspension expression that is well-formed under
  the criterion in this paper: (a)~the expression is also well-formed 
  under the earlier criterion and (b)~if the expression is an
  environment $e$, then $lev(e) \geq ind(e)$ and if $i > j$ then
  $ind_i(e) \geq ind_j(e)$. These properties must be shown
  simultaneously: the induction hypothesis pertaining to (b) is needed
  for establishing (a) and we need to know that the expression is
  well-formed in the earlier sense in order to establish (b). The
  details are straightforward once these observations are made and
  hence we omit them here. The lemma is an immediate consequence of
  property (a).
\end{proof}

The final difference between the two versions of the suspension
calculus is in the treatment of the composition of two
environments. In the earlier presentation, the outer environment is
distributed eagerly over the elements of the inner one. This is done
by a rule of the form
\begin{tabbing}
\qquad\=\kill
\>$\menv{et :: e_1, nl, ol, e_2} \ra \menvt{et,nl, ol, e_2} ::
\menv{e_1,nl,ol,e_2}$,
\end{tabbing}
where $\menvt{et,nl,ol,e_2}$ represents an augmentation to the syntax
of environment items for encoding the effect of transforming $et$ by
the relevant substitutions in $e_2$. The older version of
the calculus has rules relating to expressions of the form
$\menvt{et,nl,ol,e_2}$ that facilitate the pruning of $e_2$ down to a
part that really affects $et$ and the subsequent generation of a
suspension that captures its influence on the term component. By
contrast, the present rendition of the calculus calculates the effect
of $e_2$ on $et :: e_1$ by first pruning $e_2$ down to a relevant part
based on $et$ and only later distributing the refined environment to
$e_1$.  

It follows naturally from the observations made above that the rules
(m2), (m5) and (m6) do not appear in the original rendition of the
suspension calculus. However, based on the discussions already in
\cite{NW98tcs}, it can be seen that each of these rules is admissible to
the earlier version in the sense that their left and right hand sides
can be rewritten to a common form in that 
setting. We can, in fact, make the following observation, a detailed
proof of which appears in \cite{Gacek06msthesis}: 

\begin{lemma}\label{lem:rm-new-original}
Let $x_1$ and $x_2$ be suspension expressions such that $x_1
\readmerge x_2$. Assume further that $x_2$ is in $\onereadmerge$-normal
form. Then $x_1$ also rewrites to $x_2$ by virtue of the reading and
merging rules in \cite{NW98tcs}.
\end{lemma}

Our focus up to this point has been on arguing that the suspension
calculus as described here is a subsystem of sorts of the original
presentation. It is important, of course, to also address the issue of
why such a ``subsystem'' is of interest. There are several reasons for
this, all arising out of the modified treatment of substitution
composition. First, this treatment is a considerably simplified one
and can, as a consequence, be used directly in practical
applications. Second, it rectifies a problem with the original
calculus that prevented certain interesting logical analyses over
terms from being formulated: it is, for instance possible to describe
a type assignment system now for terms \cite{Gacek06msthesis},
something that was difficult to do with the original suspension
calculus. Finally, this change is crucial to our ability to describe
formal correspondences of the suspension calculus with other explicit
substitution calculi later in this paper.

While there may be justifications for the modified suspension
calculus, there is also a question about its adequacy. It is evident
that this version can still treat substitutions explicitly and that it
possesses the important capability of composing such substitutions. In
the next section we see also that properties such as confluence and
the ability to simulate the usual notion of beta reduction over lambda
terms are preserved, thus settling any concern over adequacy.

\subsection{Permitting Meta Variables In Suspension Terms}\label{ssec:metavars}

The syntax of suspension expressions does not presently allow for
instantiatable variables. Such variables, also referred to
as {\it meta variables}, are often used within lambda terms in
situations such as those of higher-order theorem proving and
symbolic manipulation of higher-order objects. In the former context,
these variables arise naturally in attempts to prove 
existential statements: such proofs involve choosing
instantiations for existential quantifiers and meta variables provide
a means for delaying actual choices till there is enough
information for determining what they should be. In the latter
context, instantiatable variables are instrumental in realizing
structure recognition capabilities relative to the use of higher-order
abstract syntax based representations of constructs whose structures
involve binding notions. For example, consider 
the first-order formula $\forall x ((p\ x) \lor (q\ x))$. Using an
abstraction to capture the binding content of the quantifier, this
formula can be rendered into the lambda term $(all\app \lambdadb (or\app
(p \app \#1)\app (q\app \#1)))$, where {\it all} and {\it or}
are constants chosen to encode universal quantification and
disjunction in formulas. Given such representations, the lambda term 
$(all\app \lambdadb (or\app (P \app \#1)\app (Q\app \#1)))$ in which $P$
and $Q$ are meta variables serves as a pattern for recognizing
formulas that at the top-level have the structure of a disjunction
embedded within a universal quantifier. 

An important question concerning meta variables is that of how
substitutions for them are to be treated. The logically correct
interpretation of these variables requires that such substitutions
respect the notion of scope. Thus, if $X$ is an instantiatable
variable that has an occurrence within an abstraction context, the
term that is substituted for it cannot contain a bound variable that
is captured by the enclosing abstraction. This view is one that also
supports rather useful pattern matching capabilities. To understand
this, we might reconsider the ``template'' we have described above for
first-order formulas. Suppose that we want to refine this so that the
formulas recognized by it are such that the right subpart of the
disjunction does not depend on the top-level quantifier. If a
treatment of meta variables in accordance with logical principles is
used, then the following modified template achieves this purpose:
$(all \app \lambdadb (or\app (P\app \#1)\app Q)))$. The critical facet
that ensures this behaviour is that no structure that is substituted
for $Q$ can have a variable occurrence in it that is captured by the
abstraction corresponding to the quantifier.

An alternative possibility to the logical view of instantiatable
variables is to treat them as placeholders against which any
well-formed term can be grafted. This kind of ``grafting''
interpretation has been found useful in conjunction with explicit
substitution notations in, for instance, realizing a new approach to
unification in the context of lambda terms \cite{DHK00}.  The
well-known procedure due to Huet \citeyear{Huet75} calculates unifiers
incrementally and requires the construction of a complicated term, the
contraction of beta redexes and the calculation of their substitution
effects all for the sole purpose of percolating dependency information
to places where they can be used in later computation steps. By
allowing meta variables to be substituted for by terms with variable
occurrences that can be captured by enclosing abstractions, the
dependencies can be transmitted by a much simpler process. Of course,
treating instantiatable variables in this ``graftable'' way seems
contradictory to their logical interpretation and also appears to fly
in the face of pattern matching applications. However, a
reconciliation is possible: variables can be interpreted initially in
a logical way but then surrounded in an explicit substitution context
so that a subsequent grafting treatment does not violate the required
logical constraints. Thus, consider again the term $(all \app
\lambdadb (or \app (P\app \#1) \app Q))$. This term may be transformed
into $(all \app \lambdadb (or \app (\env{P',0,1,nil}\app \#1)\app
\env{Q',0,1,nil}))$. By identifying $P$ and $Q$ with the terms
$\env{P',0,1,nil}$ and $\env{Q',0,1,nil}$, we insulate substitutions
for them from a dependence on the external abstraction even under a
grafting interpretation of $P'$ and $Q'$.

Either of the discussed views of meta variables can be built into the 
suspension notation. Towards this end, we first modify the syntax for
terms to the following:
\begin{tabbing}
\qquad\=$t$\=\quad::=\quad\=\kill
\>$t$\>\quad::=\quad\>$v\ \vert\ c \ \vert \ \#i\ \vert\ (t\app t)\ \vert\
(\lambdadb t)\ \vert\ \env{t, n, n, e}$,
\end{tabbing}
where $v$ represents the category of instantiatable variables. If we
interpret these variables in the logical way, then they cannot be 
affected by substitutions generated by $\beta$-contractions. To
support this view, therefore, we add the following to our reading rules:
\begin{tabbing}
\qquad (r7)\qquad\=\kill
\qquad (r7)\> $\env{v,ol,nl,e} \ra v$, if $v$ is a meta variable.
\end{tabbing}
If, on the other hand,  the grafting interpretation is chosen, then
this rule is not acceptable and the original rewriting system, in
fact, remains unchanged.

The choice of interpretation impact on the properties of the calculus
in different ways. Under the logical view, meta variables behave like
constants in that they may be replaced only by closed terms; this fact
is explicitly manifest in the similarity of rule (r7) to (r1). Thus,
all the properties of the calculus that includes them are already
manifest in the subsystem described in Section~\ref{ssec:syntax}. The
situation is more intricate under the grafting view. For example,
consider the term $((\lambdadb ((\lambdadb X)\app
t_1))\app t_2)$ in which $X$ is an instantiatable variable and $t_1$
and $t_2$ are terms in $\onereadmerge$-normal form. This term can be
rewritten to 
\begin{tabbing}
\qquad\=\kill
\> $\env{\env{X,1,0,(t_1,0)::nil},1,0,(t_2,0)::nil}$
\end{tabbing}
and also to 
\begin{tabbing}
\qquad\=\kill
\>
$\env{\env{X,2,1,(\#1,1)::(t_2,0)::nil},1,0,(\env{t_1,1,0,(t_2,0)::nil},0)::nil}$,
\end{tabbing}
amongst other terms. It is easy to see that these terms cannot now be
rewritten to a common form using only the reading and $(\beta_s)$
rules. The merging rules are essential to this ability. As we see 
in Section~\ref{sec:properties}, these also suffice for this
purpose. 

We assume henceforth that the suspension calculus includes meta
variables and that these are implicitly accorded the grafting
interpretation. For reasons already mentioned, it is easy to see that
the properties we establish for the resulting calculus will hold also
under the logical interpretation.

\section{Properties of the Suspension Calculus}\label{sec:properties}

We now consider the coherence of the suspension calculus. Suspensions
and the associated reading and merging rules are intended mainly to
provide control and variability over substitution relative to the
lambda calculus. In keeping with the finite nature of the substitution
process, we would expect the reduction relations defined by these
rules to be always terminating. We show this to be the case in the
first subsection. There are evidently choices to be made in the
application of the reading and merging rules. Regardless of how these
choices are made, it is important that we produce the same normal
form.  We show that this confluence property holds in the second
subsection below. We then digress briefly to establish an interesting
structural property of the suspension calculus which relates two
different methods for encoding the renumbering of bound variables;
this property is used in the next section in relating the suspension
calculus to the $\lambda\sigma$-calculus. Finally, we prove that confluence
continues to hold when the ($\beta_s$) rule is added to the collection
and that this full system is also capable of simulating beta reduction
over de Bruijn terms.

\subsection{Strong Normalizability for Substitution
  Reductions}\label{ssec:termination-rm} 

There are two steps to our argument that any sequence of rewritings
based on the reading and merging rules must terminate. First we
identify a collection of first-order terms over which we define a
well-founded ordering using a variant of recursive path orderings 
\cite{Dershowitz82,ferreira94wellfoundedness}. We then describe a
translation from suspension expressions to this collection of terms
that is such that each of the relevant rewrite rules produces a 
smaller term relative to the defined order. The desired conclusion
follows from these facts. 

The terms that are intended to capture the essence of suspension
expressions vis-a-vis termination are constructed using the
following (infinite) vocabulary: the 0-ary function symbol {\it
*}, the unary function symbol {\it lam}, and the binary function
symbols {\it app}, {\it cons} and, for each positive
number $i$, $s_i$. We denote this collection of terms by $\cal
T$. We assume the following partial ordering $\sqsupset$ on the
signature underlying $\cal T$: $s_i \sqsupset s_j$ if $i > j$ and,
for every $i$, $s_i \sqsupset {\it app}$, $s_i \sqsupset {\it
lam}$, $s_i \sqsupset {\it cons}$ and $s_i \sqsupset {\it
*}$. This ordering is now extended to the collection of terms.

\begin{defn}\label{def:termorder}
The relation $\succ$ on $\cal T$ is inductively defined by the
following property: Let $s = f(s_1,\ldots,s_m)$ and $t =
g(t_1,\ldots,t_n)$; both $s$ and $t$ may be {\it *}, \ie, the number
of arguments for either term may be $0$. Then $s \succ t$ if 
\begin{enumerate}
\item $f = g$ (in which case $n = m$), $(s_1, \ldots, s_n) \succ_{lex}
  (t_1,\ldots,t_n)$, and, $s \succ t_i$ for all $i$ such that $1 \leq
  i \leq n$, or
\item $f \sqsupset g$ and $s \succ t_i$ for all $i$ such that $1 \leq
  i \leq n$, or
\item $s_i = t$ or $s_i \succ t$ for some $i$ such that $1 \leq i \leq
  m$.
\end{enumerate}
Here $\succ_{lex}$ denotes the lexicographic ordering induced by
$\succ$. 
\end{defn}

In the terminology of \cite{ferreira94wellfoundedness}, $\succ$ is an
instance of a recursive path ordering based on $\sqsupset$. It is
easily seen that $\sqsupset$ is a well-founded ordering on the
signature underlying $\cal T$. The results in
\cite{ferreira94wellfoundedness} then imply the following: 

\begin{lemma}\label{lem:succ-wellfounded}
$\succ$ is a well-founded partial order on $\cal T$.
\end{lemma}

We now consider the translation from suspension expressions to $\cal
T$. The critical part of this mapping is the treatment of
expressions of the form $\env{t,ol,nl,e}$ and
$\menv{e_1,nl,ol,e_2}$. Our translation ignores the embedding level
components of these expressions and transforms them into terms whose
top-level function symbol is $s_i$ where $i$ is a coarse measure
of the remaining substitution work. In estimating this effort in a
sufficiently fine-grained way relative to an abstraction, it is
necessary to take cognizance of the following fact: rule (r6) creates
a ``dummy'' substitution for the bound variable that is then adjusted
by generating a ``renumbering'' suspension using rule (r3). To account
for this additional work, we define a family of measures that
relativizes the complexity of an expression to the number of enclosing
suspensions. In calculating this quantity it is important to observe
that the substitution via rule (r3) of a term in an environment
results in it being embedded in an additional suspension. We quantify
the maximum such ``internal embedding'' below and then use this in
estimating the substitution effort. In these definitions, {\it max} is
the function that picks the larger of its two integer arguments. 

\begin{defn}\label{def:intembedding} 
The measure $\mu$ that estimates the
  internal embedding potential of a suspension expression is defined as
  follows:
\begin{enumerate}
\item For a term $t$, $\mu(t)$ is $0$ if $t$ is a constant, a meta
  variable or a de Bruijn index, $\mu(s)$ if $t$ is $(\lambdadb
  s)$, ${\it  max}(\mu(s_1),\mu(s_2))$ if $t$ is $(s_1\app
  s_2)$, and  $\mu(s) + \mu(e) + 1$ if $t$ is
  $\env{s,ol,nl,e}$.  

\item For an environment $e$, $\mu(e)$ is $0$ if $e$ is nil,
  ${\it max}(\mu(s),\mu(e_1))$ if $e$ is $(s,l) :: e_1$ and
  $\mu(e_1) + \mu(e_2) + 1$ if $e$ is $\menv{e_1,nl,ol,e_2}$.
\end{enumerate}
\end{defn}

\begin{defn}\label{def:suspsize}
The measures $\eta_i$ on terms and environments for each natural
number $i$ are defined simultaneously by recursion as follows:
\begin{enumerate}
\item For a term $t$, $\eta_i(t)$ is $1$ if $t$ is a constant, a meta
  variable or a de Bruijn index, $\eta_i(s) + 1$ if $t$ is $(\lambdadb
  s)$, ${\it  max}(\eta_i(s_1),\eta_i(s_2)) + 1$ if $t$ is $(s_1\app
  s_2)$,and  $\eta_{i+1}(s) + \eta_{i+1+\mu(s)}(e) + 1$ if $t$ is
  $\env{s,ol,nl,e}$.  

\item For an environment $e$, $\eta_i(e)$ is $0$ if $e$ is nil,
  ${\it max}(\eta_i(s),\eta_i(e_1))$ if $e$ is $(s,l) :: e_1$ and
  $\eta_{i+1}(e_1) + \eta_{i+1+\mu(e_1)}(e_2) + 1$ if $e$ is
  $\menv{e_1,nl,ol,e_2}$.
\end{enumerate}
\end{defn}

The measure $\eta_0$ is meaningfully used only relative to
suspensions. In this context, it estimates, in a sense, the maximum
effort along any one path in the substitution process rather than the
cumulative effort.

\begin{defn}\label{def:susptoessence}
The translation $\ess$ of suspension expressions to $\cal
T$ is defined as follows:
\begin{enumerate}
\item For a term $t$, $\ess(t)$ is {\it *} if $t$ is a
  constant a meta variable or a de Bruijn index,
  ${\it app}(\ess(t_1),\ess(t_2))$ if $t$ is $(t_1\app t_2)$,
  ${\it lam}(\ess(t'))$ if $t$ is $(\lambdadb t')$ and
  $s_i(\ess(t'),\ess(e'))$ where $i = \eta_0(t)$ if $t$ is
  $\env{t',ol,nl,e'}$. 

\item For an environment $e$, $\ess(e)$ is {\it *} if $e$ is
  {\it nil}, ${\it cons}(\ess(t'),\ess(e'))$ if $e$ is
  $(t',l) :: e'$ and $s_i(\ess(e_1), \ess(e_2))$ where $i =
  \eta_0(e)$ if $e$ is $\menv{e_1,nl,ol,e_2}$.
\end{enumerate}
\end{defn}

We are now in a position to prove the strong normalizability of the
substitution reduction relations.

\begin{theorem}\label{th:rmtermination}
Every rewriting sequence based on the reading and merging rules
terminates.
\end{theorem}

\begin{proof}
A tedious but straightforward inspection of each of the reading and
merging rules verifies the following: If $l \ra r$ is an instance of
these rules, then $\ess(l) \succ \ess(r)$, $\mu(l) \geq \mu(r)$, and,
for every natural number $i$, $\eta_i(l) \geq \eta_i(r)$.
Definition~\ref{def:termorder} ensures that $\succ$ is monotonic, \ie,
if $v$ results from $u$ by the replacement of a subpart $x$ by $y$
such that $x \succ y$, then $u \succ v$. Further, it is easily seen
that if $x$ and $y$ are both either terms or environments such that
$\mu(x) \geq \mu(y)$ and $\eta_i(x) \geq \eta_i(y)$ for each natural
number $i$ and if $v$ is obtained from $u$ by substituting $y$ for
$x$, then $\eta_i(u) \geq \eta_i(v)$ for each natural number $i$. From
these observations it follows easily that if $t_1 \onereadmerge t_2$ then
$\ess(t_1) \succ \ess(t_2)$. The theorem is now a consequence of 
Lemma~\ref{lem:succ-wellfounded}.
\end{proof}

As an interesting side note, we observe that the termination proof
presented here has been formally verified using the {\it Coq} proof
assistant \cite{Gacek06coq}.

\subsection{Confluence for the Substitution
  Calculus}\label{ssec:confluence-rm}  

Theorem~\ref{th:rmtermination} assures us that every suspension
expression has a $\onereadmerge$-normal form. From observations in
Section~\ref{sec:notation} it follows therefore that every suspension
term can be reduced to a de Bruijn term and every environment can be
rewritten to one in a simple form using the reading and merging
rules. We now desire to show that these normal forms are unique for
any given expression. This would immediately be the case if we have
the property of confluence, \ie, if for any $s$, $u$ and $v$ such that
$s \readmerge u$ and $s \readmerge v$ we know that there must be a $t$
such that $u \readmerge t$ and $v \readmerge t$. A well-known result,
proved, for instance, in \cite{Huet80}, is that confluence follows
from a weaker property known as local confluence for a reduction
relation that is terminating. In our context this translates to it
being sufficient to show for any suspension expression $s$ that if $s
\onereadmerge u$ and $s \onereadmerge v$ then there must be an
expression $t$ such that $u \readmerge t$ and $v \readmerge t$. The
usual method for proving local confluence for a rewrite system is to
consider the different interfering ways in which pair of rules can be
applied to a given term and to show that a common term can be produced
in each of these cases. We use this approach in proving local
confluence for the reading and merging rules here. The most involved
part of the argument concerns the interference of rule (m1) with
itself. We discuss this situation first and then use our analysis in
proving the main result.

\subsubsection{An associativity property for environment composition}
\label{sssec:assoc}

The expression $\env{\env{\env{t, ol_1, nl_1, e_1}, ol_2, nl_2, e_2},
  ol_3, nl_3, e_3}$
can be transformed into a form corresponding to the
term $t$ under a substitution represented by a single environment in
two different ways by using rule (m1). The composite environments in
the two cases are given by the expressions 
\begin{tabbing}
\qquad\=\kill
\>$\menv{\menv{e_1, nl_1, ol_2, e_2}, nl_2 + (\monus{nl_1}{ol_2}),
  ol_3, e_3}$
\end{tabbing}
and 
\begin{tabbing}
\qquad\=\kill
\>$\menv{e_1, nl_1, ol_2 + (\monus{ol_3}{nl_2}), \menv{e_2,
    nl_2, ol_3, e_3}}$.
\end{tabbing}
Conceptually, these environments correspond to first composing $e_1$
and $e_2$ and then composing the result with $e_3$ or, alternatively,
to composing $e_1$ with the result of composing $e_2$ with
$e_3$. An important requirement for local confluence is that these two
environments can be made to converge to a common form, \ie,
environment composition must, in a sense, be associative. We show this
to be the case here. The argument we provide is inductive on 
the structures of the three environments and has the following broad
outline: Based on the specific context, we consider the simplification of
one of the two environments by relevant reading and merging rules. We
then show that the other expression can also be rewritten, possibly
by using the same rules, either to the same expression as the first or
to an expression that is amenable to the use of the induction hypothesis.

We begin by noting some properties of the reading and merging rules
that are useful in filling out the details of the proof. The first of
these relates to the second environment displayed above and has
the following content: At some point in the reduction of this
expression, it becomes possible to apply the rules relevant to
evaluating the composition of $e_2$ and $e_3$. Applying these rules
immediately does not limit the normal forms that can be produced. This
observation is contained in the next two lemmas. 

\begin{lemma}\label{lem:prune-back} 
Let $A$ be the environment $\menv{e_1, nl_1, ol_1, \menv{e_2, nl_2,
ol_3, e_3}}$ where $e_3$ is a simple environment and $e_2$ is of
the form $(t_2,n_2) :: e'_2$. Further, for any positive number $i$ such
that $i \leq nl_2 - n_2$ and $i \leq ol_3$, let $B$ be the environment
\begin{tabbing}
\qquad\=\kill
\>$\menv{e_1, nl_1, ol_1, \menv{e_2, nl_2 - i, ol_3 - i,
    e_3\{i\}}}$.
\end{tabbing}
If $A \readmerge C$ for any simple environment $C$ then also $B
\readmerge C$.
\end{lemma}

\begin{proof} It suffices to verify the claim when $i = 1$; an easy
  induction on $i$ then extends the result to the cases where $i >
  1$. For the case of $i = 1$, the argument is by induction on the
  length of the reduction sequence from $A$ to $C$ with the
  essential part being a consideration of the first 
  rule used. The details are straightforward and hence
  omitted. 
\end{proof}

\begin{lemma}\label{lem:unfold-back} 
Let $A$ be the environment $\menv{e_1, nl_1, ol_1, \menv{e_2, nl_2,
ol_3, e_3}}$ where $e_2$ and $e_3$ are environments of the form
$(t_2,nl_2) :: e'_2$ and $(t_3,n_3) :: e'_3$, respectively. Further,
let $B$ be the environment 
\begin{tabbing}
\qquad\=\kill
\>$\menv{e_1, nl_1, ol_1, (\env{t_2,ol_3,n_3,e_3}, n_3 +
  (\monus{nl_2}{ol_3}))::\menv{e'_2,nl_2, ol_3, e_3}}$.
\end{tabbing}
If $A \readmerge C$ for any simple environment $C$ then also $B
\readmerge C$.
\end{lemma}

\begin{proof}
The proof is again by induction on the length of the reduction
sequence from $A$ to $C$. The first rule in this sequence either
produces $B$, in which case the lemma follows immediately, or it can
be used on $B$ (perhaps at more than one place) to produce a form that
is amenable to the application of the induction hypothesis. 
\end{proof}

In evaluating the composition of $e_2$ and $e_3$, it may be the case
that some part of $e_3$ is inconsequential. The last observation that
we need is that this part can be ``pruned'' immediately in calculating
the composition of the combination of $e_1$ and $e_2$ with $e_3$. The
following lemma is consequential in establishing this fact. 

\begin{lemma}\label{lem:remvac}
Let $A$ be the environment $\menv{e_1,nl_1,ol_2,e_2}$ where
$e_2$ is a simple environment.
\begin{enumerate}
\item If $ol_2 \leq nl_1 - {\it lev}(e_1)$ then $A$ reduces to any
  simple environment that $e_1$ reduces to.
\item For any positive number $i$ such that $i \leq nl_1 - {\it
  lev}(e_1)$ and $i \leq ol_2$, $A$ reduces to any simple environment
  that $\menv{e_1,nl_1 - i, ol_2 - i, e_2\{i\}}$ reduces to.
\end{enumerate}
\end{lemma}

\begin{proof} Let $e_1$ be reducible to the simple environment
  $e'_1$. Then we may transform $A$ to the form
  $\menv{e'_1,nl_1,ol_2,e_2}$. Recalling that the level of an
  environment is never increased by rewriting, we have that ${\it
  lev}(e'_1) \leq {\it lev}(e_1)$. From this it follows that
  $A$ can be rewritten to $e'_1$ using rules (m5) and (m2) if $ol_2
  \leq nl_1 - {\it lev}(e_1)$. This establishes the first part of the
  lemma.  

The second part is nontrivial only if $nl_1 - {\it lev}(e_1)$ and
  $ol_2$ are both nonzero. Suppose this to be the case and let $B$ be
  $\menv{e_1,nl_1-1,ol_2-1,e_2\{1\}}$. The desired result follows by
  an induction on $i$ if we can show that $A$ can be rewritten to any
  simple environment that $B$ reduces to. We do this by an induction
  on the length of the reduction sequence from $B$ to the simple
  environment. This sequence must evidently be of length at least
  one. If a proper subpart of $B$ is rewritten by the first rule in
  this sequence, then the same rule can be applied to $A$ as well and
  the induction hypothesis easily yields the desired conclusion. If
  $B$ is rewritten by one of the rules (m3)-(m6), then it must be the
  case that $A \onereadmerge B$ via either rule (m4) or (m5) from
  which the claim follows immediately. Finally, if $B$ is
  rewritten using rule (m2), then $ol_2 \leq nl_1 - {\it
  lev}(e_1)$. The second part of the lemma is now a consequence of the
  first part.
\end{proof}

We now prove the associativity property for environment
composition: 

\begin{lemma}
\label{lem:assoc}
Let $A$ and $B$ be environments of the form
\begin{tabbing}
\qquad\=\kill
\>$\menv{\menv{e_1, nl_1, ol_2, e_2}, nl_2 + (\monus{nl_1}{ol_2}),
  ol_3, e_3} $
\end{tabbing}
and
\begin{tabbing}
\qquad\=\kill
\>$\menv{e_1, nl_1, ol_2 + (\monus{ol_3}{nl_2}), \menv{e_2, nl_2,
    ol_3, e_3}}$,
\end{tabbing}
respectively. Then there is a simple environment $C$ such that $A
\readmerge C$ and $B \readmerge C$.
\end{lemma}

\begin{proof}
We assume that $e_1$, $e_2$ and $e_3$ are simple environments; if this
is not the case at the outset, then we may rewrite them to such
a form in both $A$ and $B$ before commencing the proof we provide. Our
argument is now based on an induction on the structure of $e_3$ with
possibly further inductions on the structures of $e_2$ and $e_1$. 

\medskip

\noindent {\it Base case for first induction.} When $e_3$ is {\it
  nil}, the lemma is seen to be true by observing that both $A$ and
  $B$ rewrite to $\menv{e_1,nl_1,ol_2,e_2}$ by virtue of rule (m2). 

\medskip

\noindent {\it Inductive step for first induction.} Let $e_3 =
(t_3,n_3) :: e'_3$. We now proceed by an induction on the structure
of $e_2$. 

\smallskip

\noindent {\it Base case for second induction.} When $e_2$ is {\it
  nil}, it can be seen that, by virtue of rules (m2), (m3) and either
  (m4) or (m5), $A$ and $B$ reduce to $\menv{e_1,nl_1,ol_3   - nl_2,
  e_3\{nl_2\}}$ when $ol_3 > nl_2$ and to $e_1$ otherwise. The
  truth of the lemma follows immediately from this.

\smallskip

\noindent {\it Inductive step for second induction.} Let $e_2 = (t_2,n_2) ::
e'_2$. We consider first the situation where $nl_1 >
lev(e_1)$. Suppose further that $ol_3 \leq (nl_2 - n_2)$. Using rules
(m5) and (m2), we see then that 
\begin{tabbing}
\qquad\=\kill
\>$B \readmerge \menv{e_1,nl_1,ol_2,e_2}$.
\end{tabbing}
We also note that $ol_3 \leq (nl_2 + (\monus{nl_1}{ol_2})) -
lev(\menv{e_1,nl_1,ol_2,e_2})$ in this case. Lemma~\ref{lem:remvac}
assures us now that $A$ can be rewritten to any simple
environment that $\menv{e_1,nl_1,ol_2,e_2}$ reduces to and thereby
verifies the lemma in this case.

It is possible, of course, that $ol_3 > (nl_2 - n_2)$. Here we see
that   
\begin{tabbing}
\qquad\=\qquad\qquad\qquad\=\kill
\>$B \readmerge \lmenv e_1,nl_1 - 1, ol_2 + (\monus{ol_3}{nl_2}) - 1,$\\
\>\>$\menv{e'_2,n_2,ol_3 - (nl_2 - n_2),e_3\{nl_2 - n_2\}}\rmenv$.
\end{tabbing}
using rules (m5) and (m6). Using rule (m5), we also have that
\begin{tabbing}
\qquad\=\kill
\>$A \readmerge \menv{\menv{e_1,nl_1 - 1, ol_2 - 1, e'_2},nl_2 +
  (\monus{nl_1}{ol_2}),ol_3,e_3}$.
\end{tabbing}
Invoking the induction hypothesis, it follows that $A$ and 
\begin{tabbing}
\qquad\=\kill
\>$\menv{e_1,nl_1 - 1, ol_2 + (\monus{ol_3}{nl_2}) -
  1,\menv{e'_2,nl_2,ol_3,e_3}}$ 
\end{tabbing}
reduce to a common simple environment. By Lemma~\ref{lem:prune-back}
it follows that $B$ must also reduce to this environment.

The only remaining situation to consider, then, is that when $nl_1 = 
lev(e_1)$. For this case we need the last induction, that on the
structure of $e_1$.

\smallskip

\noindent {\it Base case for final induction.} If $e_1$ is {\it nil},
  then $nl_1$ must be $0$. It follows easily that both $A$ and $B$
  reduce to $\menv{e_2,nl_2,ol_3,e_3}$ and that the lemma must
  therefore be true.

\smallskip

\noindent {\it Inductive step for final induction.} Here $e_1$ must be of the
form $(t_1,nl_1) :: e'_1$. We dispense first with the
situation where $n_2 < nl_2$. In this case, by rule (m5)
\begin{tabbing}
\qquad\=\kill
\>$B \readmerge
\menv{e_1,nl_1,ol_2+(\monus{ol_3}{nl_2}),\menv{e_2,nl_2 - 1, ol_3 - 1,
    e'_3}}$.
\end{tabbing}
By the induction hypothesis used relative to $e'_3$, $B$ and the
expression 
\begin{tabbing}
\qquad\=\kill
\>$\menv{\menv{e_1,nl_1,ol_2,e_2},nl_2 + (\monus{nl_1}{ol_2}) - 1,ol_3
  - 1, e'_3}$
\end{tabbing}
must reduce to a common simple environment. By Lemma~\ref{lem:remvac},
$A$ must also reduce to this environment. 

Thus, it only remains for us to consider the situation in which $n_2 =
nl_2$. In this case by using rule (m1) twice we may transform $A$ to
the expression $A_h :: A_t$ where 
\begin{tabbing}
\qquad\=\kill
\>$A_h = (\env{\env{t_1,ol_2,n_2,e_2},ol_3,n_3,e_3}, n_3 + (\monus{(nl_2 +
  (\monus{nl_1}{ol_2}))}{ol_3}))$
\end{tabbing}
and
\begin{tabbing}
\qquad\=\kill
\>$A_t = \menv{\menv{e_1', nl_1, ol_2, e_2}, nl_2 +
  (\monus{nl_1}{ol_2}), ol_3, e_3}$.
\end{tabbing}
Similarly, $B$ may be rewritten to the expression $B_h :: B_t$ where  
\begin{tabbing}
\qquad\=$B_h\ =\ $\=$($\=\qquad\qquad\=\kill
\>$B_h =$\>$($\>$\lenv t_1, ol_2 + (\monus{ol_3}{nl_2}), n_3 +
(\monus{nl_2}{ol_3}),$ \\
\>\>\>\>$(\env{t_2, ol_3, n_3, e_3}, n_3 +
  (\monus{nl_2}{ol_3}))::\menv{e_2',nl_2,ol_3,e_3} \renv,$\\
\>\>\>$n_3 + (\monus{nl_2}{ol_3}) + (\monus{nl_1}{(ol_2 +
  (\monus{ol_3}{nl_2}))}))$
\end{tabbing}
and
\begin{tabbing}
\qquad\=\qquad\qquad\qquad\=\kill
\> $B_t = \lmenv e_1', nl_1, ol_2 + (\monus{ol_3}{nl_2}),$\\
\>\>$(\env{t_2,ol_3,n_3,e_3}, n_3 + (\monus{nl_2}{ol_3}))::\menv{e_2',
    nl_2, ol_3, e_3}\rmenv$.
\end{tabbing}
Now, using straightforward arithmetic identities, it can be seen that
the ``index'' components of $A_h$ and $B_h$ are equal. Further, the
term component of $A_h$ can be rewritten to a form identical to the
term component of $B_h$ by using the rules (m1) and (m6). Finally, by
virtue of the induction hypothesis, it follows that $A_t$ and the
expression 
\begin{tabbing}
\qquad\=\kill
\>$\menv{e_1', nl_1, ol_2 +
(\monus{ol_3}{nl_2}), \menv{e_2, nl_2, ol_3, e_3}}$
\end{tabbing}
reduce to a common simple environment. Lemma~\ref{lem:unfold-back}
allows us to conclude that $B_t$ can also be rewritten to this
expression. Putting all these observations together it is seen that
$A$ and $B$ can be reduced to a common simple environment in this case
as well.
\end{proof}

\subsubsection{Uniqueness of Substitution Normal Forms}

We can now show that $\onereadmerge$ is a locally confluent reduction
relation. 

\begin{lemma}\label{lem:localconfl}
For any expressions $s$, $u$ and $v$ such that $s \onereadmerge u$ and
$s\onereadmerge v$ there must be an expression $t$ such that $u
\readmerge t$ and $v \readmerge t$.
\end{lemma}

\begin{proof}
We recall the method of proof from \cite{Huet80}. An expression $t$
constitutes a nontrivial overlap of the rules $R_1$ and $R_2$ at a
subexpression $s$ if (a)~$t$ is an instance of the lefthand side of
$R_1$, (b)~$s$ is an instance of the lefthand side of $R_2$ and also
does not occur within the instantiation of a variable on the lefthand
side of $R_1$ when this is matched with $t$ and (c)~either $s$ is 
distinct from $t$ or $R_1$ is distinct from $R_2$. Let $r_1$ be the
expression that results from rewriting $t$ using $R_1$ and let $r_2$
result from $t$ by rewriting $s$ using $R_2$. Then the pair $\langle
r_1, r_2 \rangle$ is called the conflict pair corresponding to the
overlap in question. Relative to these notions, the lemma can be
proved by establishing the following simpler property: for every
conflict pair corresponding to the reading and merging rules, it is
the case that the two terms can be rewritten to a common form using
these rules.  

In completing this line of argument, the nontrivial overlaps that we 
have to consider are those between (m1) and each of the rules
(r1)-(r6), between (m1) and itself and between (m2) and (m3).  The  
last of these cases is easily dealt with: the two expressions
constituting the conflict pair are identical, both being {\it
  nil}. The overlap between (m1) and itself occurs over a term of the
form
$\env{\env{\env{t,ol_1,nl_1,e_1},ol_2,nl_2,e_2},ol_3,nl_3,e_3}$. By
using rule (m1) once more on each of the terms in the conflict pair,
these can be rewritten to expressions of the form $\env{t,ol',nl',e'}$
and $\env{t,ol'',nl'',e''}$, respectively, whence we can see that
$ol' = ol''$ and $nl'=nl''$ by simple arithmetic reasoning and that
$e'$ and $e''$ reduce to a common form using
Lemma~\ref{lem:assoc}. The overlaps  
between (m1) and the reading rules are also easily dealt with. For
instance consider the case of (m1) and (r1). Using rule (r1), the two
terms in the conflict pair can be rewritten to the same constant. The
other cases are similar even if a bit more tedious.
\end{proof}

As observed already, the main result of this subsection follows
directly from Lemma~\ref{lem:localconfl} and
Theorem~\ref{th:rmtermination}.

\begin{theorem}\label{th:rm-confluence}
The relation $\onereadmerge$ is confluent.
\end{theorem}

The uniqueness of $\onereadmerge$-normal forms is an immediate
consequence of Theorem~\ref{th:rm-confluence}. In the sequel, a
notation for referring to such forms will be useful.

\begin{defn}
The notation $\rnnorm{t}$ denotes the $\readmerge$-normal form of a
suspension expression $t$.
\end{defn}

It is easily seen that the $\onereadmerge$-normal form for a term 
that does not contain meta variables is a term that is devoid of
suspensions, \ie, a de Bruijn term. A further observation is that if
the all the environments appearing in the original term are simple,
then just the reading rules suffice in reducing it to the de Bruijn
term that is its unique $\onereadmerge$-normal form.

\subsection{An Equivalence Property Relating to Renumbering
  Substitutions}

An important role for the subcalculus for substitutions is that of
realizing the renumbering of de Bruijn indices necessitated by beta
contractions. One mechanism for controlling such renumbering is the new
embedding level in a suspension, \ie, the value chosen for $nl$ in an
expression of the form $\env{t,ol,nl,e}$. Looking at the reading rule
(r3), we see that another component that determines renumbering is the
index of an environment term, \ie, the value chosen for $n$ in an item
of the form $(t,n)$ in an environment. Now, these different mechanisms
appear in juxtaposition in an environment item of the form
$(\env{t,ol,nl,e},n)$.  We observe here that $\onereadmerge$-normal
forms are invariant under a coordinated readjustment of the
renumbering burden between the two devices in such an expression.

The permitted reapportionment is expressed formally through the
notion of similarity defined below.

\begin{defn}
\label{def:similar}
The similarity relation between (well-formed) terms and environments,
respectively, is denoted by $\sim$ and is given by the rules in
Figure \ref{fig:similar}.

\begin{figure}[t]
\begin{align*}
&\infer{t \sim t}{}
&
&\infer{e \sim e}{} \\
\\
&\infer{t_1\app t_2 \sim t_1'\app t_2'}{t_1 \sim t_1' && t_2 \sim t_2'}
&
&\infer{(t,n)::e \sim (t',n)::e'}{t \sim t' && e \sim e'} \\
\\
&\infer{\lambdadb t \sim \lambdadb t'}{t \sim t'}
&
&\infer{\menv{e_1,nl_1,ol_2,e_2} \sim \menv{e_1',nl_1,ol_2,e_2'}}{
  e_1 \sim e_1' && e_2 \sim e_2'} \\
\\
&\infer{\env{t,ol,nl,e} \sim \env{t',ol,nl,e'}}{t \sim t' && e \sim e'}
&
&\infer{(t,n) \sim (t',n)}{t \sim t'}
\end{align*}
\begin{align*}
\infer{(\env{t,ol,nl,r}, nl + k)::e \sim (\env{t',ol,nl',r'}, nl'
  + k)::e'}{t \sim t' && r \sim r' && e \sim e'}
\end{align*}
\caption{The similarity relation, $\sim$}
\label{fig:similar}
\end{figure}
\end{defn}

The property of interest is then the following:

\begin{theorem}\label{th:similarity}
If $t$ and $t'$ are terms such that $t \sim t'$, then $\rnnorm{t} =
\rnnorm{t'}$. If $e$ and $e'$ are environments such that $e \sim e'$,
then they rewrite by reading and merging rules to similar simple
environments. 
\end{theorem}
\begin{proof}
Only a sketch is provided here; a detailed proof may be found in
\cite{Gacek06msthesis}. Using the translation function from
Definition~\ref{def:susptoessence}, we define the relation $\gg$ on
suspension expressions as follows: $u \gg v$ just in case $\ess(u)
\succ \ess(v)$. Obviously $\gg$ is a well-founded partial order. It is
also easily seen that $u \gg v$ if either $v$ is a sub-expression of
$u$ or $u \onereadmerge v$. 

The argument is now an inductive one based on the ordering induced by
$\gg$ on pairs of expressions. In filling out the details, when
considering two expressions $u$ and $v$ such that $u \sim 
v$, the additional properties of $\gg$ and the induction hypothesis
allow us to assume that any similar subparts of $u$ and $v$ that are
terms are identical and that are environments are simple. We then
consider the different cases for the structures of $u$ and $v$ and
the rewriting rules that are applicable to them. The only
nontrivial case when $u$ and $v$ are terms arises when these are
suspensions to which rule (r3) is applicable and the environment parts
of these terms are similar but not identical. In this case we have 
\begin{align*}
u &= \env{\#1,ol,nl,(\env{t_r,ol_r,nl_r,r}, nl_r+k)::e} \\
&\one{(r3)} \env{\env{t_r,ol_r,nl_r,r}, 0, nl - (nl_r + k), nil} \\
&\one{(m1)} \env{t_r, ol_r, nl - (nl_r + k) + nl_r, \menv{r, nl_r, 0,
    nil}}\\
&\one{(m2)} \env{t_r, ol_r, nl - k, r}\\
\\
v &= \env{\#1,ol,nl,(\env{t_r,ol_r,nl_r',r'}, nl_r'+k)::e'} \\
&\one{(r3)} \env{\env{t_r,ol_r,nl_r',r'}, 0, nl - (nl_r' + k), nil}\\
&\one{(m1)} \env{t_r, ol_r, nl - (nl_r' + k) + nl_r', \menv{r', nl_r',
    0, nil}}\\
&\one{(m2)} \env{t_r, ol_r, nl - k, r'}
\end{align*}
By assumption, $r \sim r'$. Since $u \gg \env{t_r, ol_r, nl - k,
  r}$ and $v \gg \env{t_r, ol_r, nl - k, r'}$, the induction
  hypothesis yields the desired conclusion. For environments, the
  nontrivial cases arise when $u$ and $v$ are of a form to which the
  rules (m5) or (m6) apply. The argument here is similar albeit more
  tedious.
\end{proof}

Theorem~\ref{th:similarity} casts an interesting light on rule (m6) of
the suspension calculus. This rule has the form
\begin{tabbing}
\qquad\=\qquad\qquad\qquad\=\kill
\> $\menv{(t,n) :: e_1, n, ol_2, (s,l) :: e_2} \ra$\\
\>\>$(\env{t, ol_2, l, (s,l) :: e_2}, m) :: \menv{e_1, n, ol_2, (s,l)
  :: e_2}$
\end{tabbing}
where $m = l + (\monus{n}{ol_2})$. The righthand side of
the rule has an environment item in which both an index and a new
embedding level is chosen. Observe that a value larger than $l$ could
also be used for the new embedding level so long as the index is
correspondingly modified and it remains consistent with the context in
which the replacement is performed. Intuitively, this would correspond
to eagerly relativizing $\env{t,ol_2,l,(s,l)::e_2}$ to a context with
a larger number of enclosing abstractions and taking cognizance of
this in its subsequent substitution.

\subsection{Confluence for the Full Calculus}
\label{ssec:confluence-all}

Now we turn to the confluence of the system given by the rules in
Figure~\ref{fig:susprules} that includes the ($\beta_s$) rule in
addition to the ones for interpreting substitutions. In establishing
this property, we adopt the method used in \cite{CHL96jacm} to 
demonstrate that the $\lambda\sigma$-calculus is confluent. The
following lemma, proved in \cite{CHL96jacm}, is a critical part of the
argument. 

\begin{lemma}\label{lem:RSR} Let $\mathcal{R}$ and $\mathcal{S}$ be two
reduction relations defined on a set $X$ with $\mathcal{R}$ being
confluent and strongly normalizing and $\mathcal{S}$ satisfying the
property that for every $t$, $u$ and $v$ such that $t\,\mathcal{S}\,u$
and $t\,\mathcal{S}\,v$ there is an $s$ 
such that $u\,\mathcal{S}\,s$ and $v\,\mathcal{S}\,s$. Further suppose
that for every $t$, $u$ and $v$ such that $t\,\mathcal{S}\,u$ and 
$t\,\mathcal{R}\,v$ there is an $s$ such that $u\,\mathcal{R}^*\,s$ and
$v\,(\mathcal{R}^* \cup \mathcal{S} \cup\mathcal{R}^*)\, s$. Then the
relation $\mathcal{R}^* \cup \mathcal{S} \cup \mathcal{R}^*$ is
confluent. 
\end{lemma}

In applying this lemma, we shall utilize the parallelization of
$\onebetas$ that is defined below.

\begin{defn}The relation $\onebetaspar$ on suspension expressions is
  defined by the rules in Figure~\ref{fig:parallel}.
\end{defn}

\begin{figure}[t]
\begin{align*}
&\infer{t \to t}{}
&
&\infer{e \to e}{} \\
\\
&\infer{t_1\app t_2 \to t_1'\app t_2'}{t_1 \to t_1' && t_2 \to t_2'}
&
&\infer{(t,l)::e \to (t',l)::e'}{t \to t' && e \to e'} \\
\\
&\infer{\lambdadb t \to \lambdadb t'}{t \to t'}
&
&\infer{\menv{e_1,nl_1,ol_2,e_2} \to \menv{e_1',nl_1,ol_2,e_2'}}{
  e_1 \to e_1' && e_2 \to e_2'} \\
\\
&\infer{\env{t,ol,nl,e} \to \env{t',ol,nl,e'}}{t \to t' && e \to e'}
\\
\\
&\infer{(\lambdadb t_1)\app t_2 \to \env{t_1', 1, 0, (t_2',0)::nil}}{
  t_1 \to t_1' && t_2 \to t_2'}
\end{align*}
\caption{Rules defining $\onebetaspar$}
\label{fig:parallel}
\end{figure}

\begin{theorem}
\label{th:full-confluence}
The relation $\oneall$ is confluent.
\end{theorem}

\begin{proof}
Let $\mathcal{R}$ be $\onereadmerge$ and let $\mathcal{S}$ be 
$\onebetaspar$. We observe then that 
\begin{tabbing}
\qquad\=\kill
\>$\oneall \subseteq (\mathcal{R}^*\cup \mathcal{S}\cup \mathcal{R}^*)
\subseteq \readmergebetas$.
\end{tabbing}
Thus $(\mathcal{R}^*\cup \mathcal{S}\cup
\mathcal{R}^*)^* = \readmergebetas$ and hence $\oneall$ would be
confluent if $(\mathcal{R}^*\cup \mathcal{S}\cup \mathcal{R}^*)$ is. 

To establish the latter we use Lemma~\ref{lem:RSR}, interpreting
$\mathcal{R}$ and $\mathcal{S}$ as per the nomenclature of the
lemma. We have already seen that $\onereadmerge$ is both confluent and 
strongly normalizing. To show that if $t \onebetaspar u$ and $t
\onebetaspar v$ then there is an $s$ such that $u \onebetaspar s$ and
$v \onebetaspar s$, we argue by induction on the structure of $t$ and
by considering the rules by which $u$ and $v$ are obtained. The only
non-trivial case is that when $t$ is the term $(\lambdadb t_1)\app
t_2$, one of $u$ and $v$ is $\env{t'_1,1,0,(t'_2,0)::nil}$ and the
other is $(\lambdadb t''_1)\app t''_2$ where $t_1 \onebetaspar t'_1$,
$t_1 \onebetaspar t''_1$, $t_2 \onebetaspar t'_2$ and $t_2
\onebetaspar t''_2$. By the induction hypothesis, there exists an
$s_1$ such that $t'_1 \onebetaspar s_1$ and $t''_1 \onebetaspar s_1$
and an $s_2$ such that $t'_2 \onebetaspar s_2$ and $t''_2 \onebetaspar
s_2$. We then pick $s$ as $\env{s_1, 1,0,(s_2,0)::nil}$; obviously
$u \onebetaspar s$ and $v \onebetaspar s$.

It only remains for us to show that for any $t$, $u$ and $v$ such that
$t \onebetaspar u$ and $t \onereadmerge v$ there is an $s$ such that
$u \readmerge s$ and $v\,(\readmerge \cup \onebetaspar \cup
\readmerge)\, s$. We do this again by induction on the structure of
$t$. The argument is straightforward in all cases except perhaps when
$t$ is $\env{(\lambdadb t_1)\app t_2,ol,nl,e}$, $v$ is $\env{\lambdadb
  t_1, ol, nl, e}\app \env{t_2, ol, nl, e}$ and $u$ is
$\env{\env{t_1', 1, 0, (t_2',0)::nil}, ol, nl, e'}$ where $t_1
\onebetaspar t'_1$, $t_2 \onebetaspar t'_2$ and $e \onebetaspar
e'$. However, if we pick $s$ to be 
\begin{tabbing}
\qquad\=\kill
\>$\env{t_1', ol+1, nl, (\env{t_2',
    ol, nl, e'}, nl)::e'}$
\end{tabbing}
we can easily show that it satisfies the requirements, thus completing
the argument even in this case.  
\end{proof}

Theorem~\ref{th:full-confluence} strengthens the confluence result
established for the original suspension calculus in \cite{NW98tcs} in
that it shows that this property holds even when meta 
variables are permitted in terms. Although we have only shown this 
property to hold for the refinement of the suspension calculus
presented here, our argument can be easily adapted to the original
version. 

\subsection{Simulation of Beta Reduction}\label{ssec:simulation}

A fundamental requirement of any explicit substitution calculus 
is that it should allow for the simulation of beta
reduction in the usual $\lambda$-calculus. In framing this requirement
properly for the suspension calculus, it is necessary, first of all,
to restrict attention to the situation where meta variables do not
appear in terms. In this setting, as observed already, the lambda
calculus terms under the de Bruijn notation are exactly those
suspension terms that are devoid of suspensions. Moreover, beta
contraction, denoted by $\onebeta$, is defined as follows:

\begin{defn}\label{def:beta}
Let $t$ be a de Bruijn term and let
$s_1, s_2,s_3,\ldots$ represent an infinite sequence of
de Bruijn terms. Then the result of simultaneously substituting $s_i$
for the $i$-th free variable in $t$ for $i \geq 1$ is
denoted by $S(t; s_1, s_2, s_3,\ldots)$ and is defined
recursively as follows:
\begin{enumerate}
\item
$S(c; s_1, s_2, s_3,\ldots) = c$,
for any constant $c$, 

\item
$S(\#i; s_1, s_2, s_3,\ldots) = s_i$
for any variable reference $\#i$,

\item $S((t_1 \app t_2); s_1, s_2, s_3,\ldots) = 
(S(t_1 ; s_1, s_2, s_3,\ldots) \app S(t_2; s_1, s_2, s_3,\ldots))$,
and

\item
$S((\lambdadb t); s_1, s_2, s_3,\ldots) = 
(\lambdadb  S(t; \#1, s_1',s_2',s_3',\ldots))$
where, for $i \geq 1$, \hfill\break $s_i' = S(s_i; \#2,\#3,\#4,\ldots)$.
\end{enumerate}
Using this substitution operation, the $\beta$-contraction rule is
given by the following 
\begin{tabbing}
\quad (r11)\quad\=\kill
\>$( (\lambdadb t_1) \app  t_2) \ra S(t_1; t_2, \#1,\#2,\ldots)$. 
\end{tabbing}
A de Bruijn term $t$ is related via $\beta$-contraction to $s$ if $s$
results from $t$ by the application of this rule at an appropriate
subterm. We denote this relationship by $\onebeta$. Beta reduction is the
reflexive and transitive closure of $\onebeta$. 
\end{defn}

One part of the relationship between the suspension and lambda calculi
that may also be viewed as the soundness of the ($\beta_s$) rule is the
following: 

\begin{theorem}
Let $t$ and $s$ be suspension terms such that $t
\onebetas s$. Then $\rnnorm{t} \betared \rnnorm{s}$.
\end{theorem}
\begin{proof}
This theorem is proved for the original suspension calculus in
\cite{NW98tcs}. The result carries over to the version of the calculus
presented here by virtue of Lemma~\ref{lem:rm-new-original}.
\end{proof}

The ability of the suspension calculus to simulate beta reduction is a 
suitably stated converse to the above theorem.

\begin{theorem}
Let $t$ and $s$ be de Bruijn terms such that $t \betared s$. Then $t
\readmergebetas s$.
\end{theorem}
\begin{proof}
It has been shown in \cite{NW98tcs} for the original formulation of
the suspension calculus that if $t \betared s$ then $t \readbetas s$. 
This observation carries over to the present version since the
rules defining $\onereadbetas$ have essentially been preserved. The 
theorem obviously follows from this.
\end{proof}

\section{Comparison with Other Explicit Substitution
Calculi}\label{sec:compare}

We now survey some of the other explicit treatments of substitutions
that have been proposed and contrast them with the suspension
calculus. We restrict our attention in this study to calculi that
utilize the de Bruijn scheme for representing bound variables.  A good
approach to understanding such calculi is to characterize them based
on properties that are desired of them over and above their ability to
encode substitutions. These are three such properties in our
understanding: the ability to compose reduction substitutions,
confluence in a situation where graftable meta variables are included
and the preservation of strong normalizability for terms in the
underlying lambda calculus. The first of these properties is central
to combining substitution walks in normalization. Without it, for
instance, the reduction of the term $(\lambdadb \lambdadb t_1)\app t_2
\app t_3$ would require two separate traversals to be made over $t_1$
for the purpose of substituting $t_2$ and $t_3$ for the relevant bound
variables in it. The second property is important in developing
algorithms that exploit the grafting view of meta variables. For
example, confluence in the presence of such variables is a central
requirement in realizing a new approach to higher-order unification
\cite{DHK00}. The final property has both a theoretical and a
practical significance. At a theoretical level, it measures the
coherence of the calculus. Explicit treatments of substitution are
obtained usually by adding a terminating set of rules for carrying out
the substitutions generated by beta contractions. The non-preservation
of strong normalizability should, in this setting, be read as an
undesirable interference between different parts of the overall
rewrite system. At a practical level, this signifies that caution must
be exercised in designing normalization procedures.

Of these various properties, the one that appears to be most important
in practice is the ability to combine reduction substitutions: studies
show that it is central to the efficient implementation of reduction
\cite{liang-04-choices}, 
and, as indicated in Section~\ref{sec:notation}, it also appears to be
a natural way to realize confluence in the presence of graftable meta
variables. Unfortunately, the majority of the explicit substitution
calculi seem not to include this facility. Particular calculi
sacrifice other properties as well. The $\lambda\upsilon$-calculus
\cite{BBLR96} preserves strong normalizability but does not permit
graftable meta  variables. The $\lambda s_e$-calculus permits such
variables and is confluent even with this addition \cite{KR97}
but does not preserve strong normalizability \cite{Gui00jfp}. The
$\lambda\zeta$-calculus \cite{Munoz96} possesses both properties but
obtains confluence by effectively requiring beta redexes to be
contracted in an innermost fashion. Amongst the systems that do
not permit the combination of substitutions, the
$\lambda_{ws_o}$-calculus alone preserves strong normalizability and
realizes confluence in the presence of graftable meta variables
without artificially limiting reduction strategies \cite{DG01mscs}.

The only systems that permit the combination of reduction
substitutions are, to our knowledge, the $\lambda\sigma$-calculus
\cite{ACCL91}, the closely related $\Lambda$CCL calculus
\cite{Field90} and the suspension calculus. The first two calculi are
practically identical and, for this reason, we restrict our discussion
of them to only the $\lambda\sigma$-calculus. The suspension and the
$\lambda\sigma$-calculus both admit graftable meta variables without
losing confluence and they are similar in many other respects as
well\footnote{To be accurate in spirit as well as in detail this
statement needs a qualification: as we discuss later in the section,
the original rewrite system of the $\lambda\sigma$-calculus needs to
be extended slightly to obtain confluence in the presence of graftable
meta variables.}. However, they have two important differences. One of
these relates to the manner in which they represent substitutions. The
$\lambda\sigma$-calculus encodes these as independent entities that
can be separated from the term that they act on. This is a pleasant
property at a formal level but it also leads to inefficiencies in the
treatment of the renumbering of bound variables that is necessary when
a substitution is moved under an abstraction. The second difference
concerns the treatment of bound variables. In the
$\lambda\sigma$-calculus, these are encoded as environment
transforming operators in contrast to their representation directly as
de Bruijn indices in the suspension calculus. The former
representation is parsimonious in that rules that serve to compose
substitutions can also be used to interpret bound variables. However,
there are also disadvantages to such parsimony. It appears more
difficult, for example, to separate out rules based on purpose and,
hence, to identify simpler, yet complete, subsystems as has been done
for the suspension calculus \cite{Nad99jflp}. The ambiguity in
function also appears to play a role in the non-preservation of strong
normalizability in the $\lambda\sigma$-calculus \cite{Mellies95}:
although the status of this property for the suspension calculus is as
yet undetermined, a more focussed treatment of substitution
composition disallows the known counterexample for the
$\lambda\sigma$-calculus to be reproduced within it.

In the rest of this section we use the suspension calculus as a means
for understanding the different treatments of explicit substitutions
in more detail. We also attempt to substantiate the qualitative
comparisons that we have provided above. Our approach to doing this is
to describe translations between the suspension calculus and the other
calculi that illuminate their differing characteristics. None of the
calculi that we consider treat constants in terms and, for the sake of
consistency, we assume these are missing also in suspension terms. We
also do not include meta variables initially since these are not
present in all calculi, but we bring them into consideration later as
relevant. We divide our discussion of the other calculi into two
subsections depending on whether or not they possess an ability to
combine substitutions. As we shall see below, the calculi that do not
have a combining capability correspond substantially to the
suspension calculus without the merging rules.

\subsection{Calculi Without Substitution Composition}

We discuss three calculi under this rubric: the
$\lambda\upsilon$-calculus \cite{BBLR96}, the $\lambda s$-calculus
\cite{KR95}, and the $\lambda s_e$-calculus
\cite{KR97}. Qualitatively, these calculi provide an increasing
sequence of capabilities. When the de Bruijn representation is used
for lambda terms, the indices of externally bound variables in a term
have to be incremented when it is substituted under an
abstraction. The $\lambda\upsilon$-calculus requires such renumbering
to be carried out in separate walks for each abstraction that the term
is substituted under. The $\lambda s$-calculus improves on this
situation by permitting all the renumbering walks to be combined into
one although such a walk is still kept distinct from walks that
realize substitutions arising out of beta contractions. The $\lambda
s_e$-calculus extends the $\lambda s$-calculus by permitting graftable
meta variables.

\subsubsection{The $\lambda\upsilon$-calculus}

The syntax of this calculus comprises two categories: {\it terms},
corresponding to lambda terms possibly encoding explicit
substitutions, and {\it substitutions}.  

\begin{defn}
The terms, denoted by $a$ and $b$ and the substitutions, denoted by
$s$, of the $\lambda\upsilon$-calculus are given by the following
syntax rules:
\begin{tabbing}
\qquad\=$a$\quad\=::=\quad\=\kill
\>$a$\>::=\>$\underline{n} \ \vert \ a\app b\ \vert\
\lambdadb a\ \vert\ a[s]$\\
\>$s$\>::=\quad \> $a/\ \vert \lift(s)\ \vert\
\uparrow$
\end{tabbing}
\end{defn}

The collection of expressions described may be understood intuitively
as follows. The expression $\underline{n}$ represents the $n^{th}$ de
Bruijn index, analogously to $\#n$ in the suspension calculus. The
binary operator $\_[\_]$, referred to as a {\it closure}, introduces
explicit substitutions into terms. The expression $a/$, created using
the operator $/$ called {\it slash}, represents the substitution of $a$
for the first de Bruijn index and a shifting down of all other de
Bruijn indices. The substitution $\lift(s)$, which uses the operator
$\lift$ called {\it lift}, provides a device for pushing substitutions
underneath abstractions. Finally, the expression $\uparrow$,
called {\it shift}, represents the effect of increasing the de Bruijn
indices corresponding to externally bound variables by
one. 

The interpretations of the various syntactic devices are made explicit
by the rules in Figure~\ref{fig:lambda-upsilon-rules} that define the
$\lambda\upsilon$-calculus. The rule labelled (B) in this collection
emulates beta contraction by generating an explicit
substitution. The rest of the rules, that constitute the sub-calculus
$\upsilon$, serve to propagate such substitutions over the structure
of a lambda term and to eventually evaluate them at the bound variable 
occurrences. 

\begin{figure}[t]
\begin{tabbing}
\qquad \=(Lambda)\qquad\= $(\lambdadb a)[s] \ra \lambdadb a[\lift(s)]$
\qquad\qquad\= (FVarLift)\qquad\=\kill

\>(B)\> $(\lambdadb a)\app b \ra a[b/]$
\>(VarShift)\> $\underline{n}[\uparrow] \ra \underline{n+1}$\\[10pt]

\>(App)\> $(a\app b)[s] \ra a[s]\app b[s]$
\>(FVarLift)\> $\underline{1}[\lift(s)] \ra \underline{1}$ \\[3pt]
\>(Lambda)\> $(\lambdadb a)[s] \ra \lambdadb a[\lift(s)]$
\>(RVarLift)\> $\underline{n+1}[\lift(s)] \ra
                    \underline{n}[s][\uparrow]$\\[10pt]

\>(FVar)\> $\underline{1}[a/] \ra a$ \\[3pt]
\>(RVar)\> $\underline{n+1}[a/] \ra \underline{n}$
\end{tabbing}
\caption{Rewrite rules for the $\lambda\upsilon$-calculus}
\label{fig:lambda-upsilon-rules}
\end{figure}

In relating the suspension and the $\lambda\upsilon$-calculus it is
natural to identify the syntactic categories of terms in the two
settings and to think of environments in the former framework as
corresponding to substitutions in the latter. There is, however, an
important difference in the view of the latter two
entities. Substitutions in the $\lambda\upsilon$-calculus are
self-contained objects that carry all the information needed for
understanding them in context. In contrast, 
the interpretation of an environment requires also an associated old
and new embedding level in the suspension calculus. This intuition
underlies the following translation of $\lambda\upsilon$ to suspension
expressions.

\begin{defn}\label{def:lambda-upsilon-to-susp}
The mappings $T$ from terms in the $\lambda\upsilon$-calculus to terms
in the suspension calculus and $E$ from substitutions in the
$\lambda\upsilon$-calculus to triples consisting of two natural
numbers and a suspension environment are defined by
recursion as follows: 
\begin{enumerate}
\item For a term $t$, $T(t)$ is $\#n$ if $t$ is $\underline{n}$,
  $(T(a)\app T(b))$ if $t$ is $(a\app b)$, $\lambdadb T(a)$ if $t$ is
  $\lambdadb a$, and $\env{T(a),ol,nl,e}$ if $t$ is $a[s]$ and 
  $E(s) = (ol,nl,e)$.

\item For a substitution $s$, $E(s)$ is $(1,0,(T(a),0)::nil)$ if $s$
  is $a/$, $(0,1,nil)$ if $s$ is $\uparrow$, and
  $(ol+1,nl+1,(\#1,nl+1)::e)$ if $s$ is $\lift(s')$ and $E(s') = (ol,nl,e)$.
\end{enumerate}
\end{defn}

It is easy to see that $T(a)$ must be a well-formed suspension term
for every term $a$ in the $\lambda\upsilon$-calculus. The difference
in representation of bound variables in the two calculi is clearly
only a cosmetic one and we shall ignore it in the discussion that
follows. It is obvious then that $T$ is a translation that preserves
de Bruijn terms. It can also be easily verified is that $T$ and $E$
are one-to-one mappings. There are, however, {\it many} suspension
terms that are not the images under $T$ of any term in the
$\lambda\upsilon$-calculus: the set of substitutions that can be
encoded in the latter calculus is quite limited. There are, in fact,
only two forms that substitutions can take: $\lift(\ldots
\lift(a/)\ldots)$, corresponding to preserving the first few de Bruijn
indices, substituting $a$ (with appropriate renumbering) for the next
one and decreasing the remaining indices by one, and
$\lift(\ldots\lift(\uparrow)\ldots)$, corresponding to preserving the
first few de Bruijn indices and then incrementing the remaining ones
by one. Thus, the $\lambda\upsilon$-calculus cannot encode an
expression such as $\env{t,0,2,nil}$, where $t$ is a de Bruijn term,
directly. This expression can be represented indirectly by
$t[\uparrow][\uparrow]$ that has the suspension term
$\env{\env{t,0,1,nil},0,1,nil}$ as its image. This encoding highlights
a problem with the manner in which the $\lambda\upsilon$-calculus
treats renumbering of de Bruijn indices: incrementing by $n$ has to be
realized through $n$ separate walks that each increment by $1$. A more
drastic example of the limitations of the $\lambda\upsilon$-calculus
is that it possesses no simple way to encode the suspension term
$\env{t,1,2,(s,2)::nil}$ that corresponds to substituting $s$ for the
first de Bruijn index in $t$ and incrementing all the remaining
indices by two. Finally, we note that only simple environments appear
in terms that are in the image of $T$. This is, of course, to be
expected since the the $\lambda\upsilon$-calculus does not support the
ability to compose substitutions.

At the level of rewriting, we would expect the
$\lambda\upsilon$-calculus to translate into the subcalculus of the
suspension calculus that excludes the merging rules. This is true for
the most part: it is easily seen that if $l \ra r$ is an
instance of any 
rule in Figure~\ref{fig:lambda-upsilon-rules} other than (FVar) and
(RVarLift), then $T(l) \ra T(r)$ is an instance of either the
($\beta_s$) rule or one of the reading rules in
Figure~\ref{fig:susprules}. For the (FVar) rule, we observe first that
the $\env{t,0,0,nil} \ra t$ is an admissible rule in the suspension
calculus in the absence of graftable meta variables. Now, this fact
can be used to build a special case of (r3) into the rewrite system:
\begin{tabbing}
\quad\=(r3')\quad\=\kill
\> (r3')\> $\env{\#1,ol,0,(t,0) :: e} \ra t$
\end{tabbing}
The (FVar) rule corresponds directly to (r3') under the translation
we have described.

The situation for the (RVarLift) rule is more involved. Any term that
matches its lefthand side translates into a suspension term of the
form 
\begin{tabbing}
\qquad\=\kill
\>$\env{\#(n+1), ol+1, nl+1, (\#1,nl+1) :: e}$
\end{tabbing}
where either $e$ is $nil$, in which case $ol$ is $0$ and $nl$ is $1$,
or $e$ has a first element of the form $(t,nl)$. In the suspension
calculus, rule (r4) allows this term to be rewritten to the form
\begin{tabbing}
\qquad\=\kill
\>$\env{\#n, ol, nl+1, e}$.
\end{tabbing}
In the case that $e$ is $nil$, this suspension corresponds to
incrementing the indices for externally bound variables in a de Bruijn
term, constituted here by $\#n$, by 2. If $e$ is of the form $(t,nl)
:: e'$ on the other hand, then the suspension represents a situation
in which one or more terms are to be substituted into a context that
includes more enclosing abstractions than were present in the context
of their origin. The $\lambda\upsilon$-calculus is capable of
representing neither situation directly but can encode both indirectly
via a term that translates to 
\begin{tabbing}
\qquad\=\kill
\>$\env{\env{\#n,ol,nl,e},0,1,nil}$.
\end{tabbing}
This is, in fact, the translation of the righthand side of the
(RVarLift) rule. This term can be reduced to $\env{\#n,ol,nl+1,e}$ by
using the merging rules but represents the introduction of an extra
renumbering walk in the absence of these rules. 

The above discussion casts light on the efficiency with which
beta reduction can be realized using the two calculi considered
here. Normal forms for suspension expressions involving only simple
environments are identical whether or not the merging rules are
utilized. From this it follows easily that the normal forms produced
by the two systems must be identical. 

\subsubsection{The $\lambda s$-calculus}\label{ssec:lambdas}

The $\lambda s$-calculus also distinguishes between beta contraction
and renumbering substitutions. However, it differs from
the $\lambda\upsilon$-calculus in that it possesses a more general 
mechanism for renumbering de Bruijn indices and also has a more
concise way of recording which de Bruijn indices are actually affected
by beta contraction and renumbering substitutions. These devices are
manifest in the syntax of terms.
\begin{defn}
The terms of the $\lambda s$-calculus, denoted by $a$ and $b$, are
given by the rules
\begin{tabbing}
\qquad\=$a$\quad\=::=\quad\=\kill
\>$a$\>::=\>$ n \ \vert \ a\app b\ \vert\
\lambdadb a\ \vert\ a \sig{i} b\ \vert\ \ph{k}{i} a$
\end{tabbing}
where $n$ and $i$ range over positive integers and $k$ ranges over 
non-negative integers. 
\end{defn}
Towards understanding this syntax, we observe first that de Bruijn
terms are represented in the $\lambda s$-calculus exactly as they are
in the suspension calculus with the cosmetic difference that the
$n^{th}$ de Bruijn index is denoted directly by $n$ rather than $\#
n$. Beyond this, there are two additional kinds of expressions that
serve to make substitutions explicit. A term of the form $a \sig{i}
b$, called a {\it closure} and intended to capture a beta contraction
substitution, represents the substitution of a suitably renumbered
version of $b$ for the $i^{th}$ de Bruijn index in $a$ and a shifting
down by one of all de Bruijn indices greater than $i$ in $a$. A term
of the form $\ph{k}{i} a$, called an {\it update} and included to
treat renumbering, represents an increase by $i-1$ of all de Bruijn
indices greater than $k$. The purpose of these new kinds of
expressions becomes clear from the rewriting rules for the 
$\lambda s$-calculus that are presented in
Figure~\ref{fig:lambda-s-rules}. The $\sigma$-$generation$ rule is
the counterpart of beta contraction in this collection. The remaining
rules, referred to collectively as the $s$ rules, serve to calculate
substitutions introduced into terms by applications of the
$\sigma$-$generation$ rule.

\begin{figure}[t]
\begin{tabbing}
\qquad \=$\phi-app-transition$\qquad\=\kill
\>$\sigma$-$generation$\> $(\lambdadb a)\app b \ra a \sig{1} b$ \\[3pt]

\>$\sigma$-$\lambda$-$transition$\>
    $(\lambdadb a) \sig{i} b \ra \lambdadb (a \sig{i+1} b)$\\[3pt]

\>$\sigma$-$app$-$transition$\>
    $(a_1\app a_2) \sig{i} b \ra (a_1 \sig{i} b)\app (a_2 \sig{i} b)$\\[3pt]

\>$\sigma$-$destruction$\>
    $n \sig{i} b \ra
    \begin{cases}
      n - 1 & \text{if $n > i$} \\
      \ph{0}{i} b & \text{if $n = i$} \\
      n     & \text{if $n < i$}
    \end{cases}$\\[3pt]

\>$\varphi$-$\lambda$-$transition$\>
    $\ph{k}{i} (\lambdadb a) \ra \lambdadb (\ph{k+1}{i} a)$\\[3pt]

\>$\varphi$-$app$-$transition$\>
    $\ph{k}{i} (a_1\app a_2) \ra (\ph{k}{i} a_1)\app (\ph{k}{i} a_2)$\\[3pt]

\>$\varphi$-$destruction$\>
    $\ph{k}{i} n \ra
    \begin{cases}
      n + i - 1 & \text{if $n > k$} \\
      n     & \text{if $n \leq k$}
    \end{cases}$

\end{tabbing}
\caption{Rewrite rules for the $\lambda s$-calculus}
\label{fig:lambda-s-rules}
\end{figure}

Closures and updates can be understood as special forms of
suspensions. This relationship is made precise by the following
definition.  

\begin{defn}\label{def:lambda-s-to-susp}
The translation $T$ of terms in the $\lambda s$-calculus to suspension
terms is defined by recursion as follows: 
\begin{tabbing}
\qquad\=\kill
\> $T(t) = \begin{cases}
             \#n & \text{if $t$ is $n$} \\
             T(a)\app T(b) & \text{if $t = (a\app b)$} \\
             \lambdadb T(a) & \text{if $t = \lambdadb a$}\\
             \lenv T(a), i, i-1, (\#1,i-1) :: & \\
             \qquad(\#1,i-2) :: \ldots :: (\#1, 1) :: (T(b),0)::nil\renv &
             \text{if $t = a \sig{i} b$ and}\\
             \lenv T(a), k, k+i-1, (\#1,k+i-1) :: & \\
             \qquad (\#1,k+i-2) :: \ldots :: (\#1,i) :: nil \renv &
	     \text{if $t = \ph{k}{i} a$.} 
         \end{cases}
$
\end{tabbing}
\end{defn}

The image of the translation function $T$ is, once again, evidently 
a subset of the well-formed suspension terms. At a rewriting level,
the $\lambda s$-calculus is, in a sense, contained within that fragment
of the suspension calculus that excludes the merging rules.
Towards making this comment precise, we observe first that the
following is a derived rule of this fragment of the suspension
calculus, assuming that $e$ is a simple environment: 
\begin{tabbing}
\qquad\=\kill
\>$\env{\#n,ol,nl,e} = \begin{cases}
                        \#(n - ol + nl) & \text{if $n > ol$,}\\
                        \#(nl - l + 1) & \text{if $n \leq ol$ and $e[n] =
			  (\#1,l)$, and} \\
                        \env{t,0,nl - l,nil} & \text{otherwise,
			  assuming $e[n] = (t,l)$.}
                       \end{cases}$
\end{tabbing}
In particular, this rule embodies a sequence of applications of 
the rules (r2)-(r4) from Figure~\ref{fig:susprules}. Now, if we
augment the reading rules to also include this rule, then the following
theorem is easily proved: 
\begin{theorem}\label{thm:step-lambda-s}
If $a$ and $b$ are terms of the $\lambda s$-calculus such that $a$
rewrites to $b$ in one step using the rules in
Figure~\ref{fig:lambda-s-rules}, then then $T(a) \onereadbetas T(b)$.
\end{theorem}
Noting that de Bruijn terms are preserved under the translation, we
see then that any normalization sequence in the $\lambda s$-calculus
can be mimicked in a one-to-one fashion within this fragment of the
suspension calculus.

The comments above indicate a correspondence at a theoretical level
but they gloss over issues relevant to the practical implementation of
reduction. First, as the translation function indicates, the $\lambda
s$-calculus provides a rather succinct encoding for the substitutions
that arise when only the reading and the $\beta_s$ rules are
used. Second, the $s$ rules utilize this representation to realize
substitution rather efficiently in this context; observe, in this
regard, that the derived reading rule actually embodies a possibly
costly ``look-up'' operation that is necessary relative to the more
elaborate encoding of substitutions used in the suspension
calculus. However, this efficiency has an associated cost: closures in
the $\lambda s$-calculus represent exactly {\it one} beta contraction
substitution and, consequently, multiple such substitutions must be
effected in separate walks. By contrast, even simple environments
in the suspension calculus have the flexibility for encoding multiple
beta contraction and arbitrary renumbering substitutions. Moreover,
the merging rules are not needed in their full generality to exploit
this capability: simple to implement derived rules can be
described for this purpose \cite{Nad99jflp}. It has been observed that
the ability to combine substitutions that is supported by the more
general encoding for them leads to significantly greater efficiency in
realizing reduction in practice than does the concise
encoding facilitated by treating restricted forms of substitutions 
\cite{liang-04-choices}.

\subsubsection{The $\lambda s_e$-calculus and permutations of substitutions}

\begin{figure}
\begin{tabbing}
\qquad\=$\varphi$-$\varphi$-$transition$ 2\qquad\=
$(a \sig{i} b)\sig{j} c \ra (a \sig{j+1} c) \sig{i} (b \sig{j-i+1}
c)$\qquad\=\kill

\>$\sigma$-$\sigma$-$transition$
\>$(a \sig{i} b)\sig{j} c \ra
     (a \sig{j+1} c) \sig{i} (b \sig{j-i+1} c)$
\>if $i \leq j$\\[3pt]

\>$\sigma$-$\varphi$-$transition$ 1
\>$(\ph{k}{i} a)\sig{j} b \ra \ph{k}{i-1} a$
\>if $k < j < k + i$\\[3pt]

\>$\sigma$-$\varphi$-$transition$ 2
\>$(\ph{k}{i} a)\sig{j} b \ra \ph{k}{i}(a \sig{j-i+1} b)$
\>if $k + i \leq j$\\[3pt]

\>$\varphi$-$\sigma$-$transition$
\>$\ph{k}{i}(a\sig{j} b) \ra (\ph{k+1}{i} a)\sig{j} (\ph{k+1-j}{i} b)$
\>if $j \leq k + 1$\\[3pt]

\>$\varphi$-$\varphi$-$transition$ 1
\>$\ph{k}{i}(\ph{l}{j} a) \ra \ph{l}{j}(\ph{k+1-j}{i} a)$
\>if $l + j \leq k$\\[3pt]

\>$\varphi$-$\varphi$-$transition$ 2
\>$\ph{k}{i}(\ph{l}{j} a) \ra \ph{l}{j+i-1} a$
\>if $l \leq k < l + j$
\end{tabbing}
\caption{Additional rewrite rules for the $\lambda s_e$-calculus}
\label{fig:lambda-se-rules}
\end{figure}

The $\lambda s$-calculus and the $\lambda\upsilon$-calculus lack
confluence in the presence of graftable meta variables. In the absence
of substitution composition, the only way to regain confluence is to
permit permutations of substitutions\footnote{We note
  here that permutation and composition of substitutions are distinct
  notions although they seem sometimes to have been confused in the
  literature, \eg, see \cite{CKP03mscs}.}. In the context of the $\lambda
s$-calculus, such permutability should apply to both the closure and
the update forms of explicit substitutions. The $\lambda s_e$-calculus
adds the rules in Figure~\ref{fig:lambda-se-rules} to those already
present in the $\lambda s$-calculus in support of such
permutability. There must, of course, be some kind of directionality 
to the permitted substitution reorderings to ensure termination and
the side conditions on the new rules are intended to realize this. To
understand the use of these rules and also the restrictions on
permutations, we may consider the term $((\lambdadb ((\lambdadb X)\app 
t_1))\app t_2)$. Mimicking in the $\lambda s_e$-calculus the two
reduction paths seen for this term in Section~\ref{ssec:metavars},
we get the terms $(X \sig{1} t_1) \sig{1} t_2$ and $(X \sig{2} t_2)
\sig{1} (t_1 \sig{1} t_2)$. Notice now that the
$\sigma$-$\sigma$-$transition$ rule is applicable only to the first of
these terms. Thus, intuitively, this rule permits the permutation only
of substitutions arising from the contraction of outer beta redexes
over those arising from contracting inner ones. The effect of carrying
out this rearrangement is to make the substitutions have the same
form in both terms, as is desired. 

The $\lambda s_e$-calculus has been shown to have an adequate mix of
permutation rules to ensure confluence in the presence of meta
variables \cite{KR97}. From the discussion of the
$\sigma$-$\sigma$-$transition$ rule it might appear that it also
restricts these rules sufficiently to preserve strong
normalizability. Unfortunately, this is not the case: it has been
shown that interactions between closures and updatings can give rise
to nontermination even when the starting point is a lambda term that
can be simply typed \cite{Gui00jfp}. The $\lambda_{ws}$-calculus
\cite{DG01mscs} provides a remedy to this situation by extending the
syntax of de Bruijn terms (and hence the normal forms produced by
reduction) to include terms with numeric labels that represent
yet-to-be-computed renumbering substitutions. 

\subsection{Calculi with Substitution Composition}
\label{ssec:withcomposition}

As we have noted, the main exemplars of this variety of treatment of
explicit substitutions are the $\lambda \sigma$- and the suspension
calculi. We discuss their relationship below. In contrast to the
earlier situations, it is now relevant to consider mappings between
these calculi in both directions.

\subsubsection{The $\lambda\sigma$-calculus}\label{sssec:lambdasigma}

The $\lambda\sigma$-calculus, like the $\lambda\upsilon$-calculus that
is derived from it, treats substitutions as independent entities that
can be meaningfully separated from the terms they act upon. Thus, its
syntax is determined by {\it terms} and {\it substitutions}. 

\begin{defn}
The following syntax rules in which $a$ and $b$ denote terms and $s$
and $t$ denote substitutions define the syntax of the
$\lambda\sigma$-calculus: 
\begin{tabbing}
\qquad\=$q$\qquad\=::=\qquad\=\kill
\>$a$\>::=\>$1 \ \vert \ a\app b\ \vert\ \lambdadb a\ \vert\ 
a[s]$\\
\>$s$\>::=\>$id\ \vert\ a \cdot s\ \vert\ s \circ t\
\vert\ \uparrow$
\end{tabbing}
\end{defn}

\noindent A term of the form $a[s]$ is called a {\it closure} and
represents the term $a$ with the substitution $s$ to be applied to
it. The substitution $id$ is the identity substitution.  The
substitution $a \cdot s$ is called {\it cons} and represents a term
$a$ to be substituted for the first de Bruijn index along with a
substitution $s$ for the remaining indices. The substitution $s \circ
t$ represents the composition of the substitution $s$ with the
substitution $t$. Finally, the substitution $\uparrow$ is called {\it
shift} and is intended to capture the increasing by $1$ of all the de
Bruijn indices corresponding to the externally bound variables in the
term it is applied to. A form of substitution that has special
significance is $\uparrow \circ\ (\uparrow \circ\ \cdots\ (\uparrow \circ
\uparrow)\cdots)$. Assuming $n$ occurrences of $\uparrow$ in the
expression, such a substitution represents an $n$-fold increment to
the de Bruijn indices of the externally bound variables in the term it
operates on. The shorthand $\uparrow^n$ is used for such an
expression and the notation is further extended by allowing
$\uparrow^0$ to denote $id$.

The reference to de Bruijn indices in the previous paragraph is
accurate in spirit but not in detail. The $\lambda\sigma$-calculus
represents abstracted variables as environment transforming operators
rather than as indices. Specifically, only the first abstracted
variable is represented directly by the index $1$: for $n > 1$, the
$n$-th such variable is represented by $1[\uparrow^{n-1}]$. When such a
term is subjected to a substitution, the {\it shift} operators
will play a role in determining the appropriate term to replace it
with, as the rules of the calculus will elucidate. It will become
clear then that composition of substitutions is essential
in this calculus even to the proper interpretation of variables bound
by abstractions.

\begin{figure}[t]
\begin{tabbing}
\qquad \=(VarCons)\quad\= $(\lambda a)[s] \ra \lambda a[1\cdot (s\ \circ
\uparrow)]$ \qquad\qquad\= (ShiftCons)\quad\= $\uparrow \circ\
(a \cdot s) \ra s$ \kill
\>(Beta)\> $(\lambda a)\app b \ra a[b\cdot id]$ \\[20pt]
\>(App)\> $(a\app b)[s] \ra a[s]\ b[s]$ 
\>(Map)\> $(a\cdot s)\circ t \ra a[t]\cdot (s\circ t)$\\[3pt]
\>(Abs)\> $(\lambda a)[s] \ra \lambda a[1\cdot (s\ \circ
\uparrow)]$
\>(Ass)\> $(s\circ t)\circ u \ra s \circ (t\circ u)$\\[3pt]
\>(VarId)\> $1[id] \ra 1$ 
\>(IdL)\> $id\circ s \ra s$ \\[3pt]
\>(VarCons)\> $1[a\cdot s] \ra a$ 
\>(ShiftId)\> $\uparrow \circ\ id \ra\ \uparrow$ \\[3pt]
\>(Clos)\> $a[s][t] \ra a[s\circ t]$
\>(ShiftCons)\> $\uparrow \circ\ (a \cdot s) \ra s$
\end{tabbing}
\caption{Rewrite rules for the $\lambda\sigma$-calculus}
\label{fig:lambda-sigma-rules}
\end{figure}

The rules that define the $\lambda\sigma$-calculus are presented in
Figure \ref{fig:lambda-sigma-rules}. In this collection, the (Beta)
rule serves to simulate beta contraction. The remaining rules, that
define the subsystem $\sigma$, are meant to propagate substitutions
generated by the (Beta) rule. The $\sigma$ rules in the left column
compute the effect of substitutions on terms. The (Clos) rule may
generate a composition of substitutions in this process that the rules
in the right column are useful in unravelling. Given two terms or two
substitutions $u$ and $v$ , we write $u \one{\lambda\sigma} v$ or $u
\one{\sigma} v$ to denote the fact that $v$ results by replacing an
appropriate subpart of $u$ using any of these rules or only one of the
$\sigma$ rules, respectively. The reflexive and transitive closure of
these relations is, as usual, denoted by $\many{\lambda\sigma}$ and
$\many{\sigma}$.

It is useful to understand the manner in which the rules of the
$\lambda\sigma$-calculus function in the task of normalizing
terms as a prelude to contrasting it with the suspension
calculus. Towards this end, consider the lambda term given by
$(\lambdadb\lambdadb((\lambdadb\lambdadb\lambdadb \#3)\app \#2))$ in
the suspension calculus. This term is encoded by
\begin{tabbing}
\qquad\=\kill
\>$(\lambdadb\lambdadb((\lambdadb\lambdadb\lambdadb 1[\uparrow^2])\app
1[\uparrow]))$
\end{tabbing}
in the $\lambda\sigma$-calculus. Applying the (Beta)
rule to the only redex in this term we get
\begin{tabbing}
\qquad\=\kill
\>$(\lambdadb\lambdadb((\lambdadb\lambdadb
1[\uparrow^2])[1[\uparrow]\cdot id]))$.
\end{tabbing}
The substitution generated by beta contraction can now be moved inside the
two abstractions using the (Abs) rule to get the term
\begin{tabbing}
\qquad\=\kill
\>$(\lambdadb\lambdadb\lambdadb\lambdadb
(1[\uparrow^2][1\cdot (1\cdot ((1[\uparrow]\cdot id)\ \circ \uparrow)\ 
\circ \uparrow)])$.
\end{tabbing}
 The substitution $(1\cdot ((1[\uparrow]\cdot id)\ \circ \uparrow)\
\circ \uparrow)$ that appears in this expression depicts the iterated 
adjustment of  substitutions as they are pushed under abstractions
in $\lambda\sigma$-calculus; by contrast, the 
suspension calculus captures the needed renumbering simply by a global
adjustment to the new embedding level. The next conceptual step in the
reduction is that of ``looking up'' the binding for the variable
represented by $1[\uparrow^2]$ in the substitution. This step
requires the possible use of (ShiftCons) to prune off an initial
portion of the substitution and an eventual use of (VarId) to select
the desired term. However, the encoding
of abstracted variables necessitates the use of the rules (Clos),
(Ass) and (Map) to prepare the situation for applying these rules. The
term that results at the end of this process is
$(\lambdadb\lambdadb\lambdadb\lambdadb 1[(\uparrow \circ \uparrow)\
  \circ \uparrow])$. 
The (Ass) rule can now be used to transform the term under all the
abstractions into the form $1[\uparrow \circ\ (\uparrow \circ
  \uparrow)]$ that is recognizable as the encoding of a de Bruijn
index.  

\subsubsection{Translating suspension expressions into
  $\lambda\sigma$-expressions} 

The non-trivial part of this mapping concerns the treatment of 
environments in the suspension calculus. Intuitively, these must
correspond to substitutions in the $\lambda\sigma$-calculus. However,
environments obtain a meaning only relative to the new embedding level
of the suspension terms they appear in. Moreover, to be well-formed,
this embedding level must be at least as large as the level of the
environment itself. Once this constraint is satisfied, the example
just considered suggests the right translation to a ``standalone''
substitution. 

\begin{defn}\label{def:susp-to-lambda-sigma}
The mappings $S$ from suspension terms to $\lambda\sigma$-terms and
$R$ from pairs constituted by a suspension environment $e$ and a
natural number $i$ such that $lev(e) \leq i$ to
$\lambda\sigma$-substitutions are defined simultaneously by  
recursion as follows:
\begin{enumerate}
\item $S(\#1) = 1$, $S(\#(n+1)) = 1[\uparrow^n]$ if $n > 0$, $S(a\app
  b) = (S(a)\app S(b))$, $S(\lambdadb a) = \lambdadb S(a)$ and 
  $S(\env{t,ol,nl,e}) = S(t)[R(e,nl)]$.

  \item $R(e,i) = \begin{cases}
                    (\ldots ((id\ \overbrace{\circ \uparrow)\ \circ
                    \uparrow) \cdots)\ \circ \uparrow}^{i\
                    \rm{occurrences\ of}\ \uparrow}  & \text{if $e =
                    nil$}\\
                    (\ldots (((S(t) \cdot R(e',n))\
  \overbrace{\circ \uparrow)\ \circ \uparrow)\ \cdots)\ \circ
                    \uparrow}^{i-n\ \rm{occurrences\ of}\ \uparrow} &
                    \text{if $e = (t,n)::e'$ and}\\
                    R(e_1, nl_1) \circ R(e_2, i -
  (\monus{nl_1}{ol_2})) & \text{if $e = \menv{e_1, nl_1, ol_2,
                    e_2}$}.
                  \end{cases}
        $
\end{enumerate}
\end{defn}

The constraint on the pairs that $R$ applies to raises a question
concerning the well-definedness of $R$, and hence also of $S$. However,
the well-formedness requirement on suspension expressions in
Definition~\ref{def:wellformed} ensures that these must be
well-defined. Another fact that is easy to verify is that these
mappings are both one-to-one; the critical observation in this regard
is that Definition~\ref{def:susp-to-lambda-sigma} is 
constructed so that $R(e,i)$ is not equal to $\uparrow^j$ for any $e$,
$i$ and $j$. Finally, we observe a correspondence also at the level of
the rewriting:

\begin{theorem}
Let $u$ and $v$ be suspension expressions such that $u \onereadmerge 
v$ ($u \oneall v$). If $u$ and $v$ are terms, then there exists a
$\lambda\sigma$-term $w$ such that $S(u) \many{\sigma} w$ and $S(v)
\many{\sigma} w$ (respectively, $S(u) \many{\lambda\sigma} w$ and $S(v)
\many{\lambda\sigma} w$). If $u$ and $v$ are environments, then for
any $i$ such that $lev(u) \leq i$, there is a $\lambda
\sigma$-substitution $w$ such that $R(u,i) \many{\sigma} w$ and
$R(v,i) \many{\sigma} w$ (respectively $R(u,i) \many{\lambda\sigma} w$
and $R(v,i) \many{\lambda\sigma} w$). 
\end{theorem}

\begin{proof}
Applications of the rules ($\beta_s$), (r5), (r6), (m1) and (m3) on
suspension expressions map directly onto applications
of (Beta), (App), (Abs), (Clos) and (IdL), respectively, on their
translations. Rule (r2) that corresponds to renumbering a de Bruijn
index translates into a sequence of uses of the (Map) and (Ass) rules
in accordance with the representation of abstracted variables in the
$\lambda\sigma$-calculus. Rule (r3) is similar to the rule
(VarCons). However, the translation of the lefthand side
must be ``prepared'' for the use of (VarCons) by a sequence of
applications of (Map) and a peculiarity of the translation of the
righthand side may require (IdL) to be used on it to produce a common
form. In a similar sense, the rules (r4), (m4) and (m5) correspond to a
``compiled form'' of (ShiftCons) and (m6) corresponds to a compiled form
of (Map). Finally, rule (m2) is similar to the use of (Ass) in
producing a normal form. 
\end{proof}

\subsubsection{Translating $\lambda\sigma$-expressions into suspension
  expressions} 

Going in the reverse direction needs a decision on the range of the
mapping for $\lambda\sigma$-substitutions. Considering a term of 
the form $a[s]$ indicates what this might be. Such a term should
translate into a suspension of the form $\env{t,ol,nl,e}$
where the triple $(ol,nl,e)$ is obtained by ``interpreting'' $s$. In
the case when every composition in $s$ has a shift as its right
operand, this triple can be arrived at in a natural way: $e$ should
reflect the substitution terms in $s$, $ol$ should be the
number of such terms and $nl$, which counts the number of
enclosing abstractions, should correspond to the length of the longest
sequence of compositions with shifts at the top level in $s$. The
intuition underlying the encoding of general substitution composition
in the suspension calculus now allows this translation to be
extended to arbitrary $\lambda\sigma$-substitutions.

\begin{defn}\label{def:lambda-sigma-to-susp}
The mapping $T$ from $\lambda\sigma$-terms to suspension terms and the
mapping $E$ from $\lambda\sigma$-substitutions to triples of an old
embedding level, a new embedding level, and a suspension environment
are defined simultaneously by recursion as follows:
\begin{enumerate}
\item $T(1) = \#1$, $T(a\app b) = (T(a)\app T(b))$, $T(\lambdadb a) =
  \lambdadb T(a)$ and $T(a[s])$ is $\#(n+1)$ if $a$ is $1$ and $s$ is
  $\uparrow^n$ for $n \geq 0$ and is $\env{T(a),ol,nl,e}$ 
  where $E(s) = (ol,nl,e)$ otherwise.

\item $E(id) = (0,0,nil)$, $E(\uparrow) = (0,1,nil)$, $E(a \cdot s) =
  (ol+1,nl,(T(a),nl)::e)$ where $E(s) = (ol,nl,e)$, and
  $E(s_1 \circ s_2)$ is $(ol_1,nl_1+1,e_1)$ if $s_2$ is $\uparrow$ and
  is $(ol_1 + (\monus{ol_2}{nl_1}), nl_2 +
  (\monus{nl_1}{ol_2}),\menv{e_1,nl_1,ol_2,e_2})$ otherwise, assuming that
  $E(s_1) = (ol_1,nl_1,e_1)$ and $E(s_2) = (ol_2,nl_2,e_2)$.  

\end{enumerate}
\end{defn}

It is easily seen that, for any term $a$ of the
$\lambda\sigma$-calculus, $T(a)$ is a well-formed suspension term. The
translation treats a term of the form $1[\uparrow^n]$ as a special
case, reflecting its interpretation as the encoding of an abstracted
variable. If this case were not singled out, the translation
would produce the term $\env{\#1,0,nl,nil}$ instead. This term can be 
rewritten to $\#(n+1)$ by the rule (r2). A similar
observation applies to the translation of $s\app \circ \uparrow$. This 
case is treated as a special one to account for the manner in which
a substitution is moved under an abstraction in the
$\lambda\sigma$-calculus. If this issue were to be ignored,
this substitution would  translate to $(ol,nl+1,\menv{e,nl,0,nil})$
instead of $(ol,nl+1,e)$, assuming that $E(s) = (ol,nl,e)$. The
environment component of the former triple rewrites 
to that of the latter by the rule (m2).

The following theorem, whose proof is trivial, is evidence of the
naturalness of our translations:

\begin{theorem}
For every suspension term $t$, $T(S(t)) = t$.
\end{theorem}

In order to state a correspondence between the rewrite systems, 
we need to extend the reduction relations on
suspension expressions to triples of the form $(ol,nl,e)$ that are the 
targets of the mapping $E$. We do this in the obvious way: a triple
$(ol,nl,e)$ is related to $(ol,nl,e')$ by a rewriting relation just in
case $e$ is related to $e'$ by that relation. 

\begin{theorem}
If $a$ and $b$ are $\lambda\sigma$-terms such that $a \one{\sigma} 
b$ ($a \one{\lambda\sigma} b$), then there is a suspension-term $u$
such that $T(a) \readmerge u$ ($T(a) \readmergebetas u$) and $T(b)
\readmerge u$ ($T(b) \readmergebetas u$). If $s$ and $t$ are
$\lambda\sigma$-substitutions such that $s \one{\sigma} t$ ($s
\one{\lambda\sigma} t$) then there exist environments $e_1$ and $e_2$
such that $E(s) \readmerge (ol,nl,e_1)$ ($E(s) \readmergebetas
(ol,nl,e_1)$), $E(t) \readmerge (ol,nl,e_2)$ ($E(t) \readmerge
(ol,nl,e_2)$) and $e_1 \sim e_2$.
\end{theorem}
\begin{proof}
The argument is by induction on the structure of
$\lambda\sigma$-expressions. Theorem~\ref{th:similarity} permits us to
focus on the situation where rewriting takes place at the root of the
expression. Also, the observations about the ``redundancy'' of the
special cases in the definitions of $T$ and $E$ allow us to ignore
them in the proof.

Now, we can observe a relationship
between several of the rules in the $\lambda\sigma$-calculus and rules in
the suspension calculus: (Beta) corresponds to ($\beta_s$), (App)
to (r5), (Abs) to (r6), (VarId) to (a special case of) (r2), (VarCons)
to (r3), (Clos) to (m1), (IdL) to (m3), (ShiftId) to (m2) and
(ShiftCons) to (m4). In some cases the 
correspondence is precise in that the translation of the lefthand side
rewrites exactly to the translation of the righthand side by the
indicated rule. However, in most cases, some ``adjustments'' using
other reading and merging rules are needed before or after the
specific rule application to account for the peculiarities of the
different calculi. 

The two rules that remain are (Map) and (Ass). The former corresponds
to (m6) but the correspondence is not quite the same as with the other
rules. Suppose  
$(a\cdot s_1)\,\circ\, s_2$  rewrites to $a[s_2]\cdot (s_1\circ
s_2)$ by this rule. Let $T(a) = t$, $E(s_1) = (ol_1,nl_1,e_1)$, and
$E(s_2) = (ol_2,nl_2,e_2)$. The index components of $E((a\cdot
s_1)\circ s_2)$ and $E(a[s_2]\cdot (s_1\circ
s_2))$ are quickly seen to be identical. The environment components 
are $\menv{(t,nl_1)::e_1,nl_1,ol_2,e_2}$ and 
$(\env{t,ol_2,nl_2,e_2},nl_2+(\monus{nl_1}{ol_2}))::\menv{e_1,nl_1,ol_2,e_2}$,
respectively. These are like the left and right sides of rule (m6)
with two differences. First, $e_2$ might not have the form
$(s,l)::e_2'$ that is needed by rule (m6). This can be ``fixed'' by
rewriting $e_2$ at the outset to such a form\footnote{For
  completeness, the case where $e_2$ reduces 
  to $nil$ must also be discussed. (Map) in this case is related to
  (r2) and the argument is easier.}. The second difference is that the
index of the first environment term on the right side uses $nl_2$
where rule (m6) uses $l$. However, this is not a problem because the
two environments are claimed only to be similar, not identical. 

Finally, turning to (Ass), we see that there is no rule in the
suspension calculus that ``simulates'' it. Rather, this rule
corresponds to a meta property of the calculus that was proved in
Lemma~\ref{lem:assoc}.
\end{proof}

\subsubsection{Meta Variables and Preservation of Strong
  Normalizability}\label{sssec:mvar-andpsn}

Our presentation of the $\lambda\sigma$-calculus is true to its
original description in \cite{ACCL91}. This rewrite system is not
confluent when the syntax of terms is extended to include graftable meta  
variables. However, straightforward additions to the rule set
suffice to regain this property \cite{CHL96jacm}; see also
\cite{DHK00} for a system closer in form to the one discussed in this
paper.  

The $\lambda\sigma$-calculus does not preserve strong normalizability
as we have already noted, although the substitution subsystem $\sigma$
is strongly normalizing. The crux of the problem is that the (Beta)
rule and the substitution rules can interact with each other to 
get a substitution to scope over its own subcomponents. To
see how this might happen, consider the following reduction sequence
adapted from \cite{Mellies95}:
\begin{tabbing}
\qquad\=\quad\=\kill
\>$((\lambdadb a')\app b')[((\lambdadb a)\app b) \cdot id]$\\
\>\>$\many{\sigma}\quad(\lambdadb (a'[1 \cdot ((((\lambdadb a)\app b)
  \cdot id)\ \circ
  \uparrow)]))\app b'[((\lambdadb a)\app b) \cdot id]$\\
\>\>$\one{Beta}\quad a'[1 \cdot ((((\lambdadb a)\app b)
  \cdot id)\ \circ
  \uparrow)][b'[((\lambdadb a)\app b) \cdot id] \cdot id]$\\
\>\>$\many{\sigma} \quad a'[b'[((\lambdadb a)\app b) \cdot id] \cdot
  ((((\lambdadb
  a)\app b) \cdot id) \circ (\uparrow \circ\ (b'[((\lambdadb a)\app b) \cdot id] \cdot id)))]$
\end{tabbing}
The substitution $(\uparrow \circ\ (b'[((\lambdadb a)\app b) \cdot id]
  \cdot id))$ that appears as a subexpression of the last term in this
  sequence would be rewritten to $id$ in a sensible progression to a
  normal form. However, it can also perversely be distributed over the
  preceding substitution using (Map) to produce the substitution
  subexpression 
\begin{tabbing}
\qquad\=\kill
\>$((\lambdadb
  a)\app b)[\uparrow \circ\ (b'[((\lambdadb a)\app b) \cdot id]
  \cdot id)] \cdot (id \circ (\uparrow \circ\ (b'[((\lambdadb a)\app
    b) \cdot id] \cdot id)))$.
\end{tabbing}
Observe here that $[((\lambdadb a)\app b) \cdot
  id]$ has become a subpart of a substitution that stands over the
  term   $((\lambdadb a)\app b)$ that originates from itself.

The preservation of strong normalizability is still an unsettled
question with regard to the suspension calculus. However, Mellies'
counterexample does not apply to this calculus because the kind of
problem situation depicted above cannot be created within it. In
particular, rule (m6) that corresponds to (Map) in the suspension
calculus ensures that only relevant portions of an external
environment are distributed over substitution terms.

\section{Conclusion}\label{sec:conc}

This paper has presented a simplified and rationalized version of the
suspension calculus. The new notation has several pleasing theoretical
and practical properties some of which have been manifest here. This
version also differs from the original presentation in that it
preserves contextual information. This characteristic has been central
to our ability to describe translations to the
$\lambda\sigma$-calculus and has also been exploited elsewhere in
defining a system for type assignment \cite{Gacek06msthesis}. This
paper has also surveyed the world of explicit substitution calculi. It
has attempted to do this in a top-down fashion, first elucidating
properties that are important for such calculi to possess and then
using these to categorize and to explain the motivations for the
different proposed systems. In the process we have also distilled a
better understanding of the capabilities of the suspension calculus.

This work can be extended in several ways. We mention two that we
think are especially important. First, like the
$\lambda\sigma$-calculus, the notation we have described here provides
the basis for incorporating new treatments of higher-order unification
that exploit graftable meta variables into practical systems. It is of
interest to actually explicate such a treatment and to evaluate its
benefits empirically. Second, the question of preservation of strong
normalizability is still an open one for this calculus. This issue
appears to be a non-trivial one to settle and an answer to it is
likely to provide significant insights into the structure of the
suspension calculus.

\begin{acks}
This work began while the second author was on a sabbatical visit to
the Protheo group at LORIA and INRIA, Nancy and the Comete and
Parsifal groups at {\'E}cole Polytechnique and INRIA, Saclay. Support
for this work has been provided by the NSF through the grant numbered
CCR-0429572; however, any opinions, findings, and conclusions or
recommendations expressed in this paper are those of the authors and
do not necessarily reflect the views of the National Science
Foundation. Gacek has also been supported by a grant from Boston
Scientific during the concluding stages of this research.
\end{acks}

\bibliography{../thisbib}
\bibliographystyle{acmtrans}

\end{document}